\documentclass[aps,twocolumn,showpacs,preprintnumbers,amsmath,amssymb,nofootinbib,superscriptaddress,showkeys]{revtex4-1}
\usepackage{graphicx}

\newcommand{\amu}{a_{\mu}}
\usepackage{color}

\begin{document}

\title{{\boldmath $\eta$} and {\boldmath $\eta^\prime$} transition form factors from rational approximants}
\author{Rafel Escribano}
\email{rescriba@ifae.es}
\affiliation{
Grup de F\'{\i}sica Te\`orica (Departament de F\'{\i}sica) and Institut de F\'{\i}sica d'Altes Energies (IFAE),
Universitat Aut\`onoma de Barcelona, E-08193 Bellaterra (Barcelona), Spain
}
\author{Pere Masjuan}
\email{masjuan@kph.uni-mainz.de (Corresponding author)}
\author{Pablo Sanchez-Puertas}
\email{sanchezp@kph.uni-mainz.de}
\affiliation{Institut f\"ur Kernphysik, Johannes Gutenberg-Universit\"at, D-55099 Mainz, Germany}
\date{\today}

\begin{abstract}
The $\eta$ and $\eta^\prime$ transition form factors in the space-like region are analyzed at low and intermediate energies
in a model-independent way through the use of rational approximants.
The slope and curvature parameters of the form factors as well as their values at zero and infinity are extracted from
experimental data.
The impact of these results on the mixing parameters of the $\eta$-$\eta^\prime$ system and
the pseudoscalar-exchange contributions to the hadronic light-by-light scattering part of the anomalous magnetic moment
$a_{\mu}$ are also discussed.
\end{abstract}



\maketitle

\section{Introduction}
\label{intro}

The pseudoscalar transition form factors (TFFs) $\gamma^*\gamma\to P$, where $P=\pi^0, \eta, \eta^\prime$ or $\eta_c$,
have attracted a lot of attention recently, both from the experimental and theoretical sides, 
since the release of the \textit{BABAR} data on the $\pi^0$-TFF in 2009 \cite{Aubert:2009mc}.
The TFF describes the effect of the strong interaction on the $\gamma^*\gamma^*\to P$ transition and 
is represented by a function $F_{P\gamma^*\gamma^*}(q_1^2,q_2^2)$ of the photon virtualities.
Measuring both virtualities from the two-photon-fusion reaction $e^+e^-\to e^+e^-P$ is still an experimental challenge,
so the common practice is to extract the TFF when one of the outgoing leptons is tagged and the other is not, that is,
the single-tag method.
The tagged lepton emits a highly off-shell photon with the momentum transfer $q_1^2\equiv -Q^2$ and is detected,
while the other, untagged, is scattered at a small angle and its momentum transfer $q_2^2$ is near zero.
The form factor extracted from the single-tag experiment is then a function of one of the virtualities:
$F_{P\gamma^*\gamma}(Q^2)\equiv F_{P\gamma^*\gamma^*}(-Q^2,0)$.  

At low-momentum transfer, the TFF can be described by the expansion
\begin{equation}
\label{eq:tffexpansion}
F_{P\gamma^*\gamma}(Q^2)=F_{P\gamma\gamma}(0)\left(1-b_P\frac{Q^2}{m_P^2}+c_{P}\frac{Q^4}{m_P^4}+\cdots\right)\ ,
\end{equation}
where $F_{P\gamma\gamma}(0)$ is the normalization,
the parameters $b_P$ and $c_P$ are the slope 
(related to the mean square radius of the meson by $b_P/m_P^2=\langle r^2\rangle/6$)
and the curvature, respectively,
and $m_P$ is the pseudoscalar meson mass.
$F_{P\gamma\gamma}(0)$ can be obtained either from the measured two-photon partial width of the meson $P$,
\begin{equation}
\label{eq:tffdecay}
|F_{P\gamma\gamma}(0)|^2=\frac{64 \pi}{(4\pi\alpha)^2}\frac{\Gamma(P\to\gamma\gamma)}{m_P^3}\ ,
\end{equation}
or, in the case of $\pi^0$, $\eta$ and $\eta^\prime$,
from the prediction of the axial anomaly in the chiral and large-$N_c$ limits of QCD.
For instance, $F_{\pi^0\gamma\gamma}(0)=1/(4\pi^2 F_\pi)$, where $F_\pi\simeq 92$ MeV is the pion decay constant.
The corresponding predictions for the $\eta$ and $\eta^\prime$ are discussed below.
Concerning the slope parameter, chiral perturbation theory (ChPT) predicts \cite{Bijnens:1988kx,Bijnens:1989jb}
$b_{\eta}=0.51$ and  $b_{\eta^\prime}=1.47$
for $\sin\theta_P=-1/3$ \cite{Ametller:1991jv},
being
$\theta_P$ the $\eta$-$\eta^\prime$ mixing angle in the octet-singlet basis defined at lowest order.
Other theoretical predictions are~\cite{Ametller:1991jv}:
$b_{\eta}=0.53$ and $b_{\eta^\prime}=1.33$, from vector meson dominance (VMD);
$b_{\eta}=0.51$ and $b_{\eta^\prime}=1.30$, from constituent-quark loops; 
$b_{\eta}=0.36$ and $b_{\eta^\prime}=2.11$, from the Brodsky-Lepage interpolation formula \cite{brodsky-lepage3}; and
$b_{\eta}=0.521(2)$ and $b_{\eta^\prime}=1.323(4)$, from resonance chiral theory \cite{Czyz:2012nq}.
More recently, the values $b_{\eta}=0.61^{+0.07}_{-0.03}$ and $b_{\eta^\prime}=1.45^{+0.17}_{-0.12}$
have been obtained from a dispersive analysis for $\eta\to\gamma\gamma^\star$ \cite{Hanhart:2013vba}.
Experimental determinations of these parameters are usually obtained after a fit to data using a normalized, 
single-pole term with an associated mass $\Lambda_P$,
\textit{i.e.,}~
\begin{equation}
\label{eq:tffpole}
F_{P\gamma^*\gamma}(Q^2)=\frac{F_{P\gamma\gamma}(0)}{1+Q^2/\Lambda_P^2}\ .
\end{equation}

At large-momentum transfer, the TFF can be calculated in the asymptotic $Q^2\rightarrow \infty$ limit at leading twist as a convolution of a perturbative hard scattering amplitude $T_H(\gamma\gamma^* \rightarrow q \bar{q} )$ and a gauge-invariant meson distribution amplitude which incorporates the nonperturbative dynamics of the QCD bound state \cite{brodsky-lepage2}. 

While the low- and large-momentum transfer regions are in principle well described by ChPT and perturbative QCD (pQCD), respectively,
it would be highly desirable to have a model-independent description of the TFFs in the whole energy range.
Unfortunately, such a description is still lacking for the $\eta$ and $\eta^\prime$
\cite{Feldmann:1997vc,Kroll:2010bf,Dorokhov:2011zf,Brodsky:2011yv,Brodsky:2011xx,Klopot:2011qq,Wu:2011gf,DiDonato:2011kr,Noguera:2011fv,Balakireva:2011wp,
Czyz:2012nq,Melikhov:2012bg,Melikhov:2012qp,Geng:2012qg,Klopot:2012hd,Kroll:2012hs}
(see also a first attempt beyond single-pole interpolation formulas in Ref.~\cite{Feldmann:1998yc}).
In Ref.~\cite{Masjuan:2012wy},
it was suggested for the $\pi^0$ case that this model-independent description can be achieved using a sequence of rational functions, the Pad\'e approximants (PAs),
to fit the experimental data.
In this way, not only the low- and large-momentum transfer predictions of ChPT and pQCD should be reproduced but also a reliable description of the intermediate-energy region would be available.
The main advantage of the method of PAs is indeed to provide the $Q^2$ dependence of the TFF over the whole space-like region in an easy, systematic and model-independent way \cite{Masjuan:2008fv,Masjuan:2012wy}. 
This is relevant, for instance, when extrapolating from the asymptotic $Q^2$ limit to the charmonium region~\cite{Rosner:2009bp}.
We also notice that for the forthcoming KLOE-2 \cite{Babusci:2011bg} and BES-III \cite{Asner:2008nq} TFFs measurements will be helpful to have a more reliable model-independent description of the whole energy range, and particularly at low energies, in order to build up a solid Monte Carlo generator for data analysis and feasibility studies.

The aim of this work is then to extend and further develop the application of PAs, already initiated in Refs.~\cite{Masjuan:2008fv,Masjuan:2012wy},
to the analysis of $\eta$ and $\eta^\prime$ TFFs taking into account $\eta$-$\eta^\prime$ mixing effects systematically.
As shown below, this analysis complements our understanding of the $\eta$-$\eta^\prime$ mixing pattern and, more important, can shed light on their relation to the anomalous magnetic moment of the muon, $a_{\mu}$, through its hadronic light-by-light scattering contribution (HLBL). Preliminary results were presented in Ref.~\cite{Czyz:2013zga}.

The paper is organized as follows.
In Sec.~\ref{sec:PSTFF}, we briefly describe the general method for extracting low-energy parameters from the TFFs using rational approximants
and then apply this method to the case of $\eta$ and $\eta^\prime$ TFFs.
In Sec.~\ref{sec:mixing}, we discuss the implications of our results for the determination of the $\eta$-$\eta^\prime$ mixing parameters.
Finally, in Sec.~\ref{sec:g2}, we analyze the possible impact of our findings on the HLBL piece of $a_{\mu}$,
with special attention to the $\eta$ and $\eta^\prime$ exchange contributions.
The conclusions are presented in Section~\ref{sec:conc}.

\section{{\boldmath $\eta$} and {\boldmath $\eta^\prime$} transition form factors at low and intermediate energies}
\label{sec:PSTFF}

In order to extract the low-energy parameters $b_P$ and $c_P$ (slope and curvature, respectively) from the available data, we use the method described in
Refs.~\cite{Masjuan:2008fv,Masjuan:2012wy}.
This method makes use of PAs as fitting functions to all the experimental data in the space-like region.
PAs are rational functions $P^N_M(Q^2)$ [ratio of a polynomial $T_N(Q^2)$ of order $N$ and a polynomial $R_M(Q^2)$ of order $M$]
constructed in such a way that they have the same Taylor expansion as the function to be approximated up to order ${\cal O} (Q^2)^{N+M+1}$ \cite{Baker}.
Since PAs are built in our case from the unknown low-energy parameters (LEPs) of the TFF, once the fit to the experimental data is done,
the reexpansion of the PAs yields the desired coefficients.
We refer the interested reader to Ref.~\cite{Masjuan:2012wy} for details on this technique.

The main feature of this method is the usage of a sequence of PAs.
In this way, one can ascribe a systematic error to the result assuming a convergent behavior of the sequence\footnote{
The convergence of the sequence can only be mathematically proven (and not assumed) for certain types of special functions 
(see Ref.~\cite{Baker} for details).}.
This systematic error is defined as the relative error between the prediction of a finite-order PA for a given parameter and the result from the exact function, becoming eventually zero.
However, since the exact function of the TFF is unknown, the assumption of convergence can only be checked against well-motivated models.
After performing such a test for several different models, one takes as the systematic error the most conservative result.
It is in this way that we consider the PAs a systematic and model-independent approach.
In App.~\ref{App}, such a test of convergence with a Regge model for the $\eta$ TFF is shown.
As soon as convergence is assumed, the largest the sequence is, the smallest the systematic error turns out to be.
In a realistic case the sequence will not be infinite since, at some given order, the additional parameters of the fitted PAs will be statistically compatible with zero.
Then, one should stop the sequence at that order
leading to the intrinsic or systematic error on the LEPs predictions explained above.
In Refs.~\cite{Masjuan:2008fv,Masjuan:2012wy}, this error was carefully studied and provided.
In accordance with this, we ascribe a conservative systematic error of the order of $5\%$ and $20\%$
for the slope and curvature parameters, respectively, to our final LEP determinations\footnote{
In Appendix~\ref{App}, the Regge model used to show convergence yields smaller systematic errors.
Thus, to be conservative, we prefer to take the results discussed in Refs.~\cite{Masjuan:2008fv,Masjuan:2012wy}.}.
Since in practice, our PA sequences are quite short (up to 5-6 elements at most),
one needs to consider several kinds of sequences with different analytical properties for better strengthening the results.
This procedure avoids problems of overfitting as well.
The choice of which type of PA sequence to be used is largely determined by the analytic properties of the function to be approximated.
As argued in Ref.~\cite{Masjuan:2012wy}, the time-like region for the $\pi^0$-TFF exhibits a predominant role of the $\rho$ meson contribution
with the excited states being much suppressed.
For the $\eta$ and $\eta^\prime$ TFF, the appropriate combination of the $\rho,\omega$ and $\phi$ mesons should play the same role through an effective single-pole dominance
as the $\rho$ on the $\pi^0$ TFF.
For that reason, a $P^L_1(Q^2)$ sequence (single-pole approximants) seems the optimal choice in the $\eta$ and $\eta^\prime$.
However, according to Ref.~\cite{brodsky-lepage2}, the pseudoscalar TFFs behave as $1/Q^2$ for $Q^2\to\infty$,
which means that, for any value of $L$, one will obtain in principle a good fit only up to a finite value of $Q^2$ but not for $Q^2\to\infty$.
Therefore, it would be desirable to incorporate this asymptotic limit information in the fits by considering also a $P^N_{N}(Q^2)$ sequence.

In the following subsection, we present and discuss the weighted averaged results for the LEPs of the $\eta$ and $\eta^\prime$ TFFs
obtained from the PA sequences mentioned above.
Since it is common to publish experimental data in the form of $Q^2 F_{\eta^{(\prime)} \gamma^*\gamma}(Q^2)$ instead of
$F_{\eta^{(\prime)} \gamma^*\gamma}(Q^2)$, we prefer to fit the first form.
We do this following a bottom-up approach.
So, we start fitting the $Q^2 F_{\eta^{(\prime)} \gamma^*\gamma}(Q^2)$ space-like data without any information at $Q^2=0$.
This means, in particular, that the mathematical $\lim_{Q^2\to 0} Q^2 F_{\eta^{(\prime)} \gamma^*\gamma}(Q^2)=0$ is not imposed but extracted from data.
In a second step, we impose such limit making use of PAs whose numerator starts at order $Q^2$, \textit{i.e.}~$T_N(0)=0$.
This will allow us to predict the value of the transition form factors at zero, and therefore the two-photon partial widths, from pure space-like data, 
as well as the slope and curvature parameters.
Finally, as a last step, we incorporate the measured two-photon partial widths in our set of data to be fitted together with the space-like data.
The former bottom-up approach should allow us to strengthen systematically our results.

\subsection{{\boldmath $\eta$} and {\boldmath $\eta^\prime$} transition form factors}
\label{etaTFF}

For both the $\eta$ and $\eta^\prime$ TFFs,
we collect the experimental data from the CELLO\footnote{
The CELLO Collaboration does not report a systematic error for each bin of data.
While for the $\eta^\prime$ case such error is $16\%$ of the total number of events
(which we translate into $32\%$ for each bin), for the $\eta$ case, only $12\%$ for the two-photon channel is reported.
Accounting for all the different systematic sources we could find in the publication,
we ascribe $12\%$ of systematic error for the hadronic $\eta$ decay which leads to $6\%$ error for the global number of events
(implying $12\%$ of systematic error for each bin).},
CLEO, L3, and \textit{BABAR} Collaborations \cite{Behrend:1990sr,Gronberg:1997fj,Acciarri:1997yx,BABAR:2011ad}.
As stated, we include in our final step the values
$\Gamma_{\eta\to\gamma\gamma}=0.516(18)$~keV
(obtained after combining the PDG average \cite{PDG2012} together with the recent KLOE-2 result \cite{Babusci:2012ik})
and $\Gamma_{\eta^\prime\to\gamma\gamma}=4.35(14)$~keV from the PDG fit \cite{PDG2012}.
Since the asymptotic values of the space-like and time-like TFFs are expected to be very similar,
we also comment on the results when including in our analysis the time-like measurements\footnote{
The time-like TFF for the $\pi^0$ at high energy is not yet available,
but it could be measured at $q^2=14.6$ GeV$^2$ by the BES-III Collaboration \cite{Asner:2008nq}.
This particular point is in the region, reaching the asymptotic limit, where the measurements from the \textit{BABAR} and Belle Collaborations
start to differ, so we encourage BES-III to measure it.}
for the $\eta$ and $\eta^\prime$ reported by the \textit{BABAR} Collaboration \cite{Aubert:2006cy}.

We start our bottom-up approach by fitting space-like data alone without including the constraint
$\lim_{Q^2\to 0} Q^2 F(Q^2)=0$.
For the $\eta$ case, the fits ``see the zero" within 2 standard deviations.
For instance, the coefficient that would be fixed to zero in case this constraint is imposed is found to be $0.059(29)$ for a fit to a $P^1_1(Q^2)$ approximant.
For the $\eta^\prime$ case, the results are better and the zero is seen within one standard deviation.
As an example, we find $-0.002(3)$ for a fit to a $P^3_1(Q^2)$ approximant.
Once this coefficient is seen to be zero, the next one is identified with $F_{\eta^\prime\gamma\gamma}(0)$
and found to be $0.38(6)$ GeV$^{-1}$, which, making use of Eq.~(\ref{eq:tffdecay}), leads to the prediction
$\Gamma_{\eta^\prime\to\gamma\gamma}=5.3(1.7)$ keV.
This fact illustrates the potentiality of the space-like data,
which ranges from $0.6$ to $35$ GeV$^2$ for the $\eta$ and from $0.06$ to $35$ GeV$^2$ for the $\eta^\prime$,
to shed light on the low-energy region of these TFFs.
\begin{figure*}[htbp]
\begin{center}
\includegraphics[width=8.5cm]{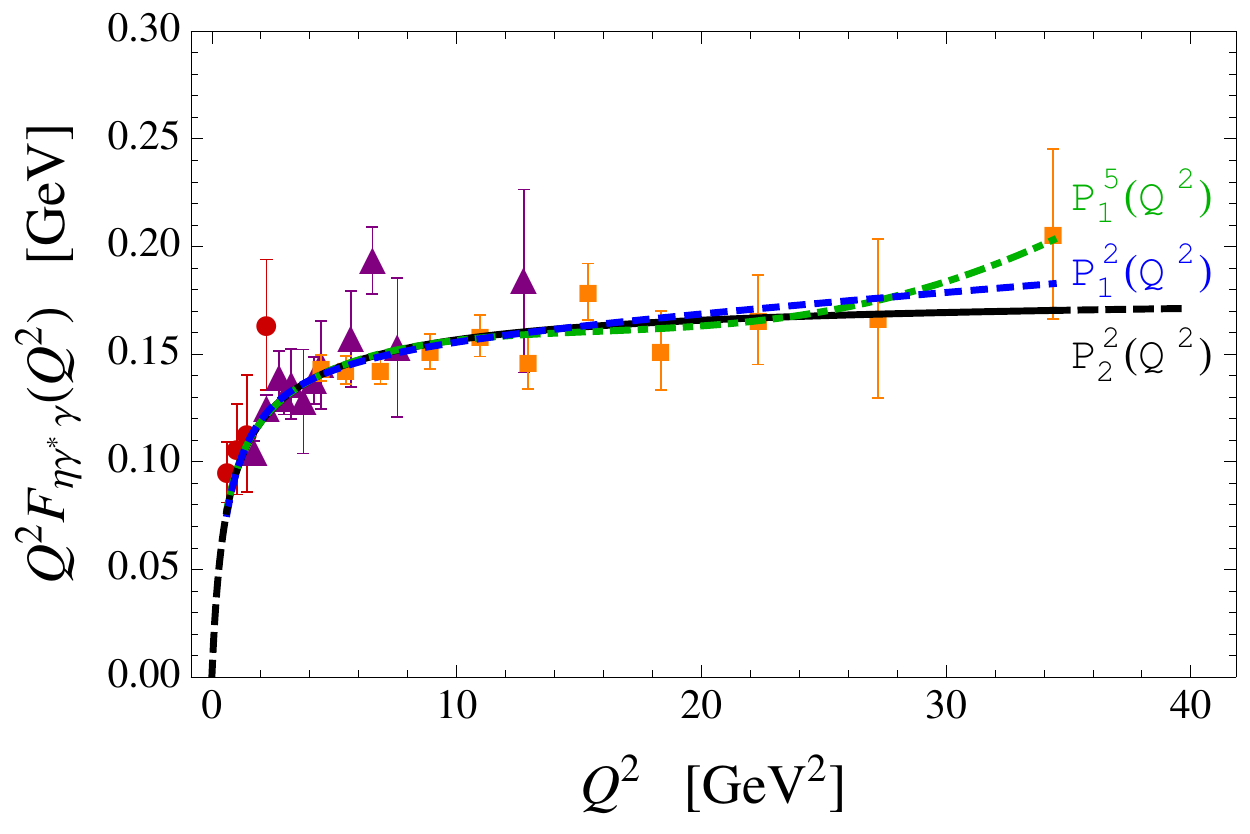}
\includegraphics[width=8.5cm]{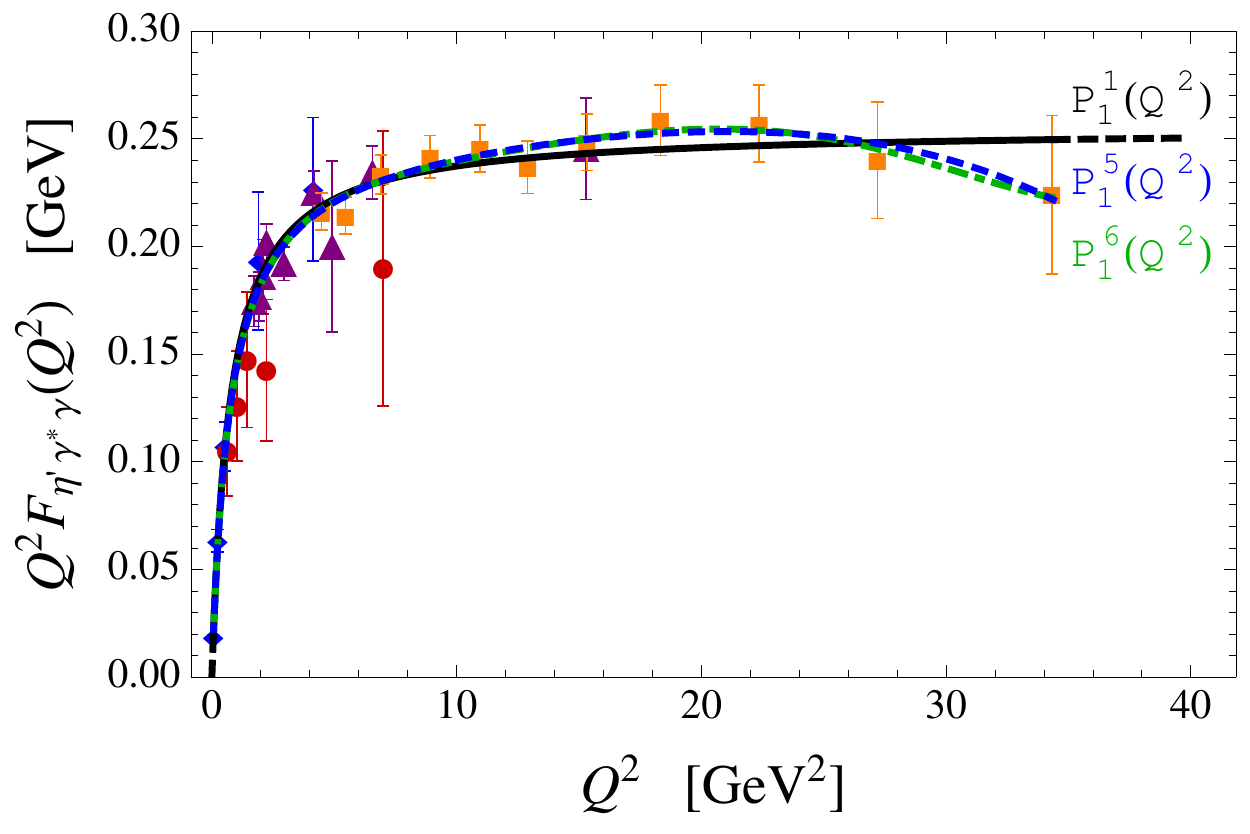}
\caption{$\eta$ (left panel) and $\eta^\prime$ (right panel) TFFs best fits.
Blue-dashed lines show our best $P^L_{1}(Q^2)$ when the measured two-photon partial decay widths \emph{are not} included in the fits,
green-dot-dashed lines show our best $P^L_{1}(Q^2)$ when the two-photon widths \emph{are} included,
and black-solid lines show our best $P^N_{N}(Q^2)$ in the latter case. 
Black-dashed lines display the extrapolation of the $P^N_{N}(Q^2)$ at $Q^2=0$ and $Q^2\to\infty$.
Experimental data points are from CELLO (red circles) \cite{Behrend:1990sr},
CLEO (purple triangles) \cite{Gronberg:1997fj},
L3 (blue diamonds) \cite{Acciarri:1997yx}, and 
{\textit{BABAR}} (orange squares) \cite{BABAR:2011ad} Collaborations.}
\label{fig:etaetap}
\end{center}
\end{figure*}
\begin{table*}[ht]
\caption{Low-energy parameters for the $\eta$ and $\eta^\prime$ TFFs obtained from the PA fits to experimental data 
\emph{without including} the measured two-photon partial decay widths.
The first column indicates the type of sequence used for the fit and $N$ is the highest order reached with that sequence.
The last row shows the weighted average result for each LEP.
We also present the quality of the fits in terms of $\chi^2$/DOF (degrees of freedom).
Errors are only statistical and symmetrized.}
\begin{center}
\begin{tabular}{|c||c|c|c|c|c||c|c|c|c|c|}
\hline
& \multicolumn{5}{c||}{$\eta$ TFF} & \multicolumn{5}{c|}{$\eta^\prime$ TFF}\\
\cline{2-11}
& $N$ & $b_{\eta}$ & $c_{\eta}$ & $F_{\eta\gamma\gamma}(0)$ GeV$^{-1}$ & $\chi^2$/DOF 
& $N$ & $b_{\eta^\prime}$ & $c_{\eta^\prime}$ & $F_{\eta^\prime\gamma\gamma}(0)$ GeV$^{-1}$ & $\chi^2$/DOF\\
\hline
$P^N_{1}(Q^2)$ & $2$ & $0.45(13)$ & $0.20(12)$ & $0.235(53)$ & $0.79$ & $5$ & $1.25(16)$ & $1.57(42)$ & $0.339(17)$ & $0.70$\\
$P^N_{N}(Q^2)$ & $1$ & $0.36(6)$ & $0.13(4)$ & $0.201(28)$ & $0.78$ & $1$ & $1.19(6)$ & $1.42(15)$ & $0.332(15)$ & $0.68$\\
\hline
Final & $ $ & $0.45(13)$ & $0.20(12)$ & $0.235(53)$ & $$ & $$ & $1.25(16)$ & $1.57(42)$ & $0.339(17)$ & $$\\
\hline
\end{tabular}
\end{center}
\label{tab:psresults}
\end{table*}
\begin{table*}[ht]
\caption{Low-energy parameters for the $\eta$ and $\eta^\prime$ TFFs obtained from the PA fits to experimental data 
\emph{including} the measured two-photon partial decay widths.
The first column indicates the type of sequence used for the fit and $N$ is the highest order reached with that sequence.
The last row shows the weighted average result for each LEP.
We also present the quality of the fits in terms of $\chi^2$/DOF.
Errors are only statistical and symmetrized.}
\begin{center}
\begin{tabular}{|c||c|c|c|c||c|c|c|c|}
\hline
& \multicolumn{4}{c||}{$\eta$ TFF} & \multicolumn{4}{c|}{$\eta^\prime$ TFF}\\
\cline{2-9}
& $N$ & $b_{\eta}$ & $c_{\eta}$ & $\chi^2$/DOF & $N$ & $b_{\eta^\prime}$ & $c_{\eta^\prime}$ & $\chi^2$/DOF\\
\hline
$P^N_{1}(Q^2)$ & $5$ & $0.58(6)$ & $0.34(8)$ & $0.80$ & $6$ & $1.30(15)$ & $1.72(47)$ & $0.70$\\ 
$P^N_{N}(Q^2)$ & $2$ & $0.66(10)$ & $0.47(15)$ & $0.77$ & $1$ & $1.23(3)$ & $1.52(7)$ & $0.67$\\
\hline
Final & $ $ & $0.60(6)$ & $0.37(10)$ & $ $ & $ $ & $1.30(15)$ & $1.72(47)$ & $ $\\
\hline
\end{tabular}
\end{center}
\label{tab:psresults2}
\end{table*}

\begin{figure*}[htbp]
\begin{center}
\includegraphics[width=8.5cm]{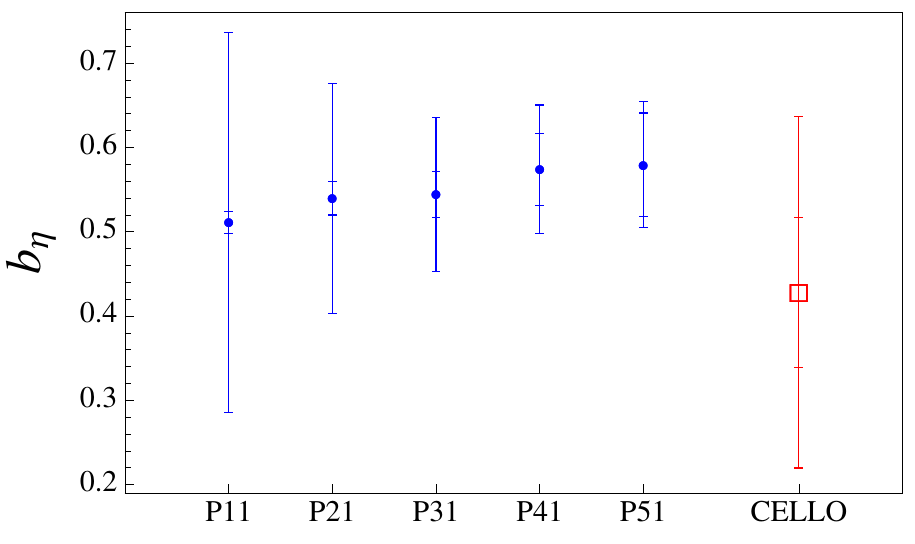}
\includegraphics[width=8.5cm]{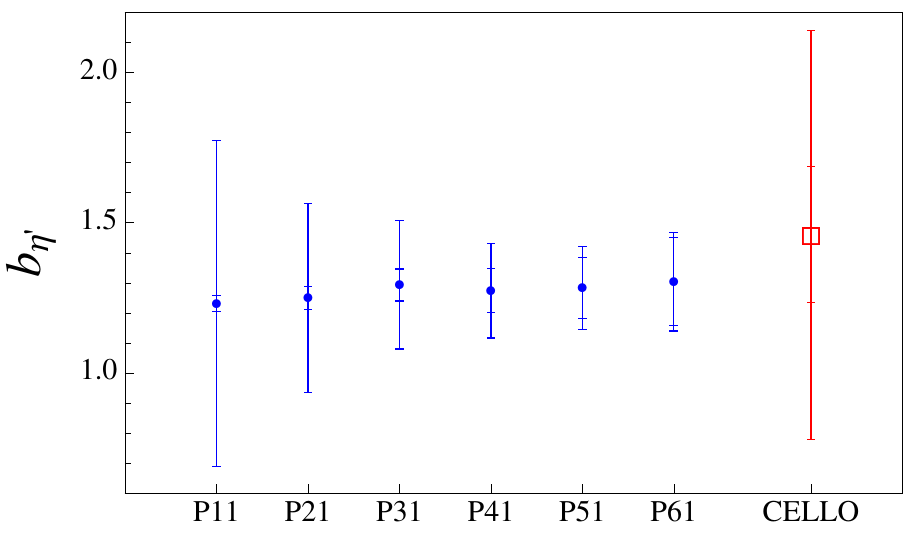}
\caption{Slope predictions for the $\eta$ (left panel) and $\eta^\prime$ (right panel) TFFs using the $P^L_1(Q^2)$ up to $L=5$ for the $\eta$ 
and $L=6$ for the $\eta^\prime$, respectively (blue circles).
The internal bands correspond to the statistical error of the different fits and the external ones are the combination of statistical and systematic errors determined as explained in the main text.
The CELLO determination is also shown for comparison (empty-red squares).}
\label{fig:slope}
\end{center}
\end{figure*}

\begin{figure*}[htbp]
\begin{center}
\includegraphics[width=8.5cm]{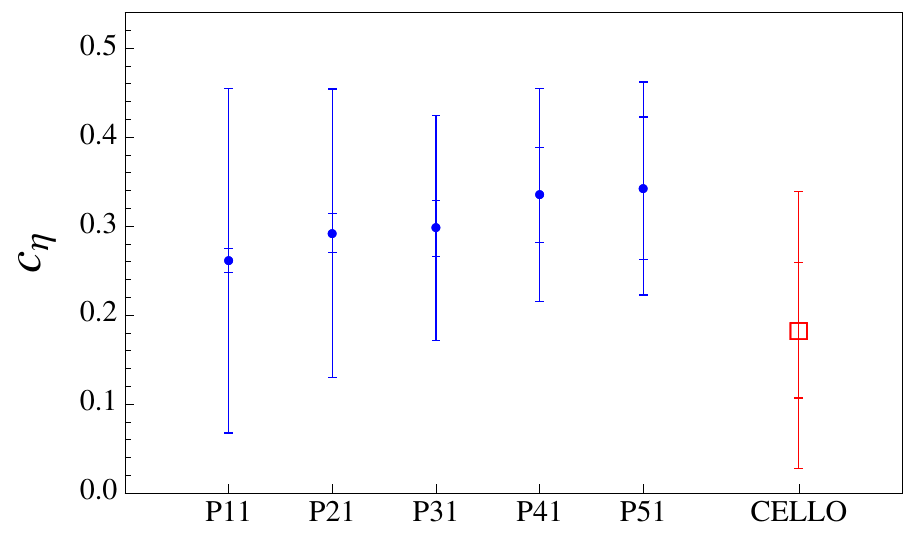}
\includegraphics[width=8.5cm]{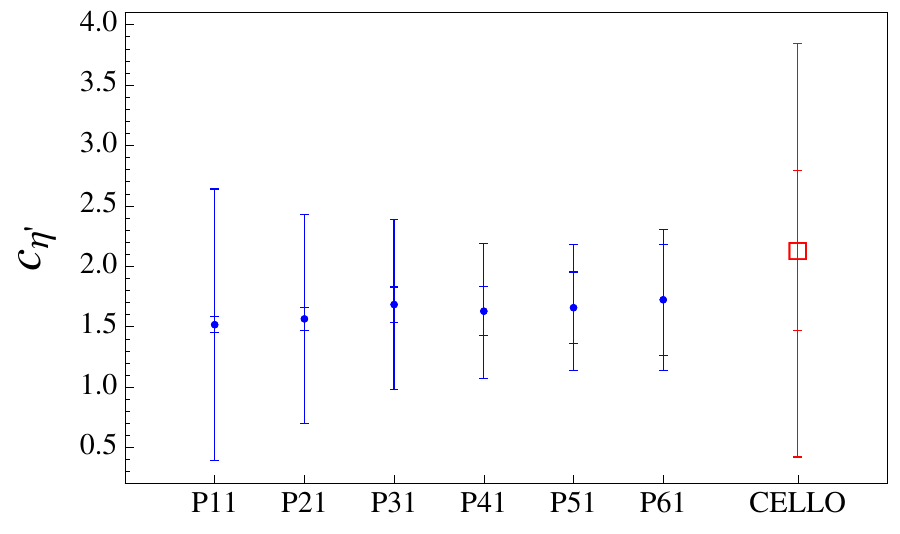}
\caption{Curvature predictions for the $\eta$ (left panel) and $\eta^\prime$ (right panel) TFFs using the $P^L_1(Q^2)$ up to $L=5$ for the $\eta$ 
and $L=6$ for the $\eta^\prime$, respectively (blue circles).
The internal bands correspond to the statistical error of the different fits and the external ones are the combination of statistical and systematic errors determined as explained in the main text.
The CELLO determination is also shown for comparison (empty-red squares).}
\label{fig:curv}
\end{center}
\end{figure*}

The next step is to include $\lim_{Q^2\to 0} Q^2 F(Q^2)=0$ into the fits, that is,
to consider PAs whose numerator starts already at order $Q^2$.
Our best fitted approximants for the $\eta$ and $\eta^\prime$ TFFs in this case are shown in Fig.~\ref{fig:etaetap} for two scenarios:
\emph{with} and \emph{without} including the two-photon partial widths in the data set.
The obtained LEPs are collected in Table \ref{tab:psresults}, $\Gamma_{\eta^{(\prime)}\to\gamma\gamma}$ not included,
and Table \ref{tab:psresults2}, $\Gamma_{\eta^{(\prime)}\to\gamma\gamma}$ included,
and shown in Fig.~\ref{fig:slope}, for the slope, and Fig.~\ref{fig:curv}, for the curvature, respectively. 
The stability observed for the LEPs with the $P^L_1(Q^2)$ is quite reassuring.
For completeness, we also include in these figures the results obtained by the CELLO Collaboration \cite{Behrend:1990sr}
using a VMD model fit.
To perform an appropriate comparison of their LEPs with our results,
we add to their determinations the same systematic error we included in ours,
which turns out to be of the order of $40\%$ following Refs.~\cite{Masjuan:2008fv,Masjuan:2012wy}.
The coefficients of the best fitted $P^L_1(Q^2)$ approximants
when the constraint at $Q^2=0$ and the experimental two-photon decay widths are included
can be found in Appendix~\ref{AppTL} for both TFFs.

All the different PAs considered so far lead to compatible results.
However, for the LEP determinations, the inclusion of the measured two-photon partial widths in the fits is crucial for two reasons:
first, smaller errors on such decays immediately yield smaller errors on the slope and curvature parameters;
second, and more important, precise two-photon partial widths allow us to reach higher PAs in our sequences, rendering smaller systematic errors.
Then, more precise measurements of such partial widths will be very welcome for extracting the LEPs
with better statistical and systematical errors.
If instead, the two-photon partial widths are not included in the analysis, these are still well determined by our fits.
Using Eq.~(\ref{eq:tffdecay}) and the fitted values for $F_{\eta^{(')}\gamma^*\gamma}(0)$ from Table \ref{tab:psresults},
we predict such partial widths to be $\Gamma_{\eta\to\gamma \gamma}=0.38(17)$ keV and
$\Gamma_{\eta^\prime\to\gamma \gamma}=4.22(42)$ keV.
These results only differ from the measured ones by $0.8$ and $0.3$ standard deviations, respectively.
We remark on the importance of the high-energy TFF experimental data to obtain large PA sequences
which in turn permit better determinations of the LEPs.

The last row in Tables \ref{tab:psresults} and \ref{tab:psresults2}
presents our final results for the LEPs obtained after a weighted average of the different determinations depending on the type of PA sequence used.
We consider the values shown in Table \ref{tab:psresults2}, when the measured two-photon partial widths are taken into account,
as the main results of this work, while the results in Table \ref{tab:psresults} are kept for comparison. 
For this reason, in the following, we only comment on the results of Table \ref{tab:psresults2}.
\begin{figure*}[htbp]
\begin{center}
	\includegraphics[width=0.45\textwidth]{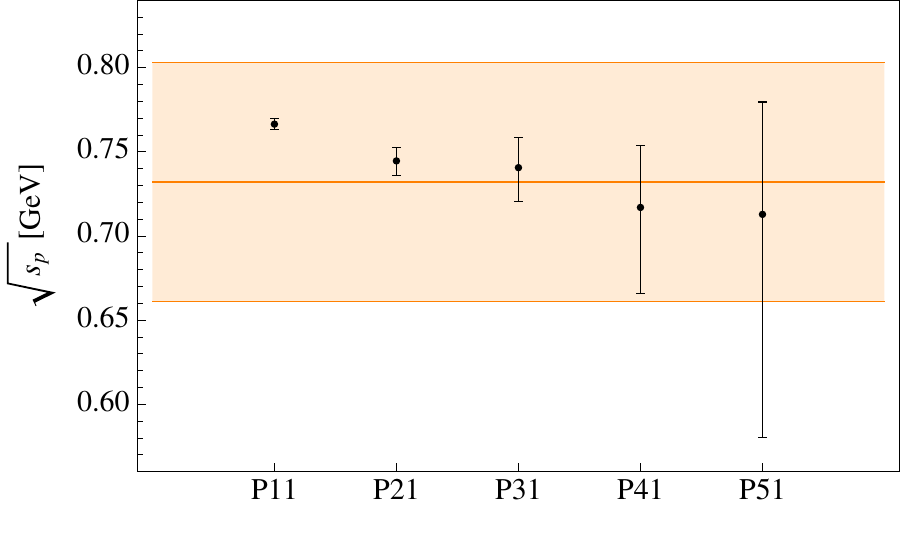}
	\includegraphics[width=0.45\textwidth]{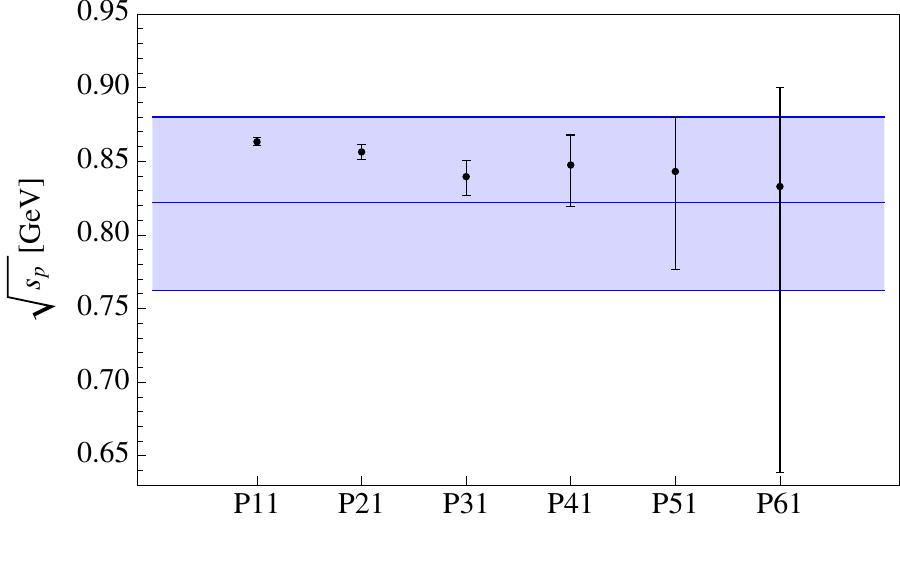}
\caption{Pole-position predictions for the $\eta$ (left panel) and $\eta^\prime$ (right panel) TFFs using the $P^L_1(Q^2)$
up to $L=5$ for the $\eta$ and $L=6$ for the $\eta^\prime$, respectively.
For comparison, we also display (orange and blue bands) the range $m_{\rm eff}\pm\Gamma_{\rm eff}/2$
corresponding to the effective VMD meson resonance evaluated using the half-width rule (see main text for details).}
\label{fig:PL1poles}
\end{center}
\end{figure*}

\begin{figure*}[tbp]
\begin{center}
	\includegraphics[width=0.49\textwidth]{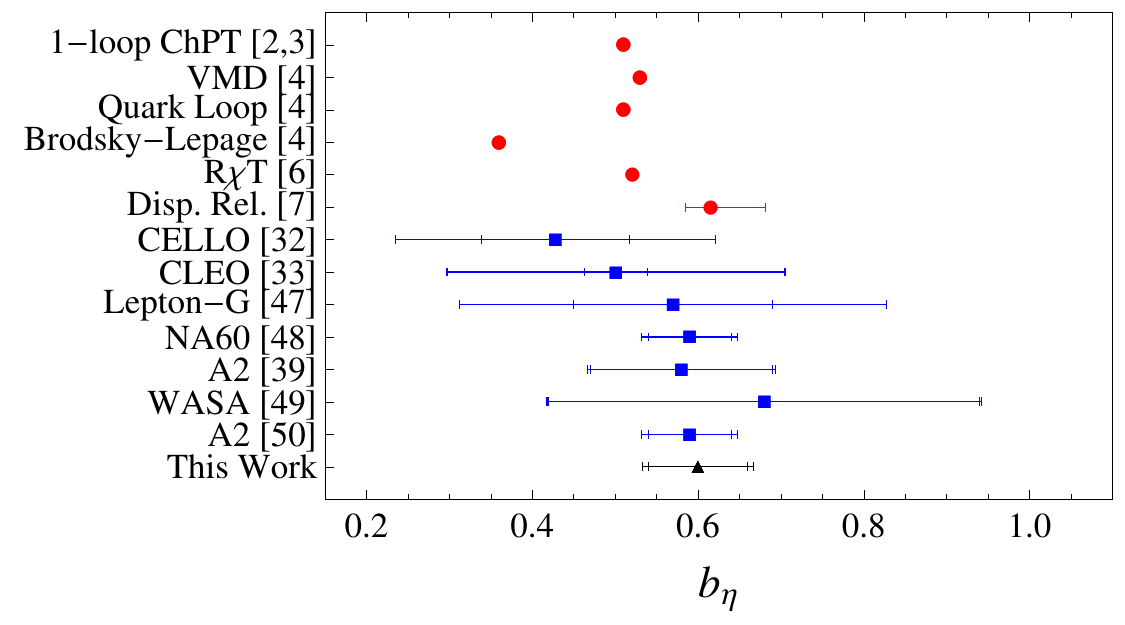}
	\includegraphics[width=0.49\textwidth]{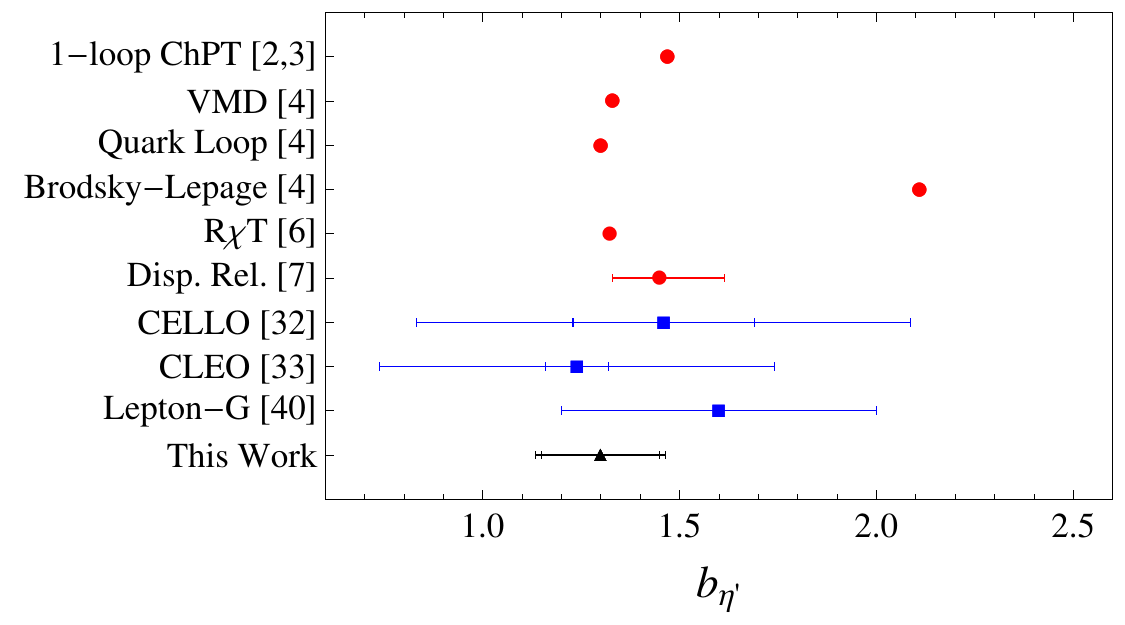}
\caption{Slope determinations for  $\eta$ (left panel) and $\eta^\prime$ (right panel) TFFs from different theoretical (red circles) and experimental (blue squares) references discussed in the text. Inner error is the statistical one and larger error is the combination of statistical and systematic errors.}
\label{fig:slopecomp}
\end{center}
\end{figure*}

For the $\eta$ and $\eta^\prime$, respectively, the $P^L_1(Q^2)$ sequence reaches $L=5$ and $L=6$ as the best approximant.
The fitted poles range $\sqrt{s_p}$=(0.71--0.77) GeV and $\sqrt{s_p}$=(0.83--0.86) GeV, as can be seen in Fig.~\ref{fig:PL1poles}.
For comparison, we also show as orange and blue bands what would correspond to the effective VMD meson resonance $m_{\rm eff}$
\cite{Landsberg:1986fd}, using $m_{\rho}=0.775$~GeV, $\Gamma_{\rho}=0.148$~GeV, $m_{\omega}=0.783$~GeV, $\Gamma_{\omega}=0.008$~GeV, $m_{\phi}=1.019$~GeV, and $\Gamma_{\phi}=0.004$~GeV.
The bands represent the range of such mass values due to the half-width rule \cite{Masjuan:2012gc,Masjuan:2012sk,Masjuan:2013xta},
\textit{i.e.}~$m_{\rm eff}\pm\Gamma_{\rm eff}/2$.
We obtain $m_{\rm eff}=0.732(71)$ GeV for the $\eta$ case and $m_{\rm eff}=0.822(58)$ GeV for the $\eta^\prime$,
with errors due to the half-width rule.
Notice that raising the poles lowers the LEPs (slope and curvature) and vice versa.
As shown, fitting space-like data does not produce an accurate determination of the resonance poles as already indicated in
Refs.~\cite{Masjuan:2007ay,Masjuan:2008fr,Masjuan:2008fv,Masjuan:2012wy}.
Thus, we do not recommend to apply this method for such determinations. 
That includes the use of VMD fits to determine the resonance parameters.
An alternative model-independent procedure of extracting these parameters using PAs can be found in Ref.~\cite{Masjuan:2013jha}.

To reproduce the asymptotic behavior of the TFFs, we have also considered the $P^N_{N}(Q^2)$ sequence
(second row in Tables \ref{tab:psresults} and \ref{tab:psresults2}).
The results obtained are in nice agreement with our previous determinations.
The best fits are shown as black-solid lines in Fig.~\ref{fig:etaetap}.
We reach $N=2$ for the $\eta$ case and $N=1$ for the $\eta^\prime$.
Since these approximants contain the correct high-energy behavior built in, they can be extrapolated up to infinity
(black-dashed lines in Fig.~\ref{fig:etaetap}) and then predict the leading $1/Q^2$ coefficient:
\begin{equation}
\label{mixinglimits}
\begin{split}
\lim_{Q^2\to\infty }Q^2F_{\eta\gamma^*\gamma}(Q^2) &=0.160(24)\ \textrm{GeV}\ ,\\
\lim_{Q^2\to\infty }Q^2F_{\eta^\prime\gamma^*\gamma}(Q^2) &=0.255(4)\ \textrm{GeV}\ .
\end{split}
\end{equation}
We emphasize once more the importance of including the measured two-photon partial widths in the fits,
that for the case of the $\eta$ TFF allows us to reach $N=2$ and then reduce the uncertainty drastically.
Otherwise, we would have remained at $N=1$ with errors 5 times larger. 

Finally, our combined weighted average results from Table \ref{tab:psresults2},
taking into account both types of sequences, give
\begin{equation}
\label{etaetapvalues}
\begin{split}
&\begin{cases}
 	&b_{\eta} = 0.60(6)_{\rm stat}(3)_{\rm sys}\\
 	&c_{\eta} = 0.37(10)_{\rm stat}(7)_{\rm sys}
 \end{cases}
\\
&\begin{cases}
	&b_{\eta^\prime} = 1.30(15)_{\rm stat}(7)_{\rm sys}\\
	&c_{\eta^\prime} = 1.72(47)_{\rm stat}(34)_{\rm sys}
 \end{cases}
\end{split}
\end{equation}
where the second error is systematic (of the order of $5\%$ and $20\%$ for $b_P$ and $c_P$, respectively).
When the spread of central values considered for the weighted averaged result is larger than the error after averaging,
we enlarge this error to cover that spread\footnote{We thank C.F.~Redmer for discussions on the average procedure.}
\cite{PDG2012}.
Equation~(\ref{etaetapvalues}) represents the main results of this work.
For the case of the $\eta^\prime$, with the $P^N_N(Q^2)$ sequence we could only reach $N=1$,
which turns out to be the first element on the $P^L_1(Q^2)$ sequence.
The first element of each sequence is the worst and should not be taken for final averaged results. 

For the $\eta$, the slope of the TFF obtained in Eq.~(\ref{etaetapvalues}) can be compared with
$b_{\eta}=0.428(89)$ from CELLO \cite{Behrend:1990sr} and $b_{\eta}=0.501(38)$ from CLEO \cite{Gronberg:1997fj}.
The TFF was also measured in the time-like region with the results
$b_{\eta}=0.57(12)$ from Lepton-G \cite{Dzhelyadin:1980kh}, 
 $b_{\eta}=0.585(51)$ from NA60 \cite{Arnaldi:2009aa}, 
$b_{\eta}=0.58(11)$ from A2 \cite{Berghauser:2011zz}, and 
$b_{\eta}=0.68(26)$ from WASA \cite{Hodana:2012rc}. Recently, the A2 Collaboration reported $b_{\eta}=0.59(5)$~\cite{Aguar-Bartolome:2013vpw}, the most precise experimental extraction up to date.
For the $\eta^\prime$, the slope in Eq.~(\ref{etaetapvalues}) can be compared with
$b_{\eta^\prime}=1.46(23)$ from CELLO \cite{Behrend:1990sr},
$b_{\eta^\prime}=1.24(8)$ from CLEO \cite{Gronberg:1997fj},
and $b_{\eta^\prime}=1.6(4)$ from the time-like analysis by the Lepton-G Collaboration (cited in Ref.~\cite{Landsberg:1986fd}).
One should notice that all the previous collaborations used a VMD model fit to extract the slopes.
In order to be consistent when comparing with our results, a systematic error of about $40\%$ should be added to the experimental determinations based on space-like data and a smaller one of about $5\%$ on the ones based on time-like data, smaller since such data are closer to $Q^2=0$~\cite{Masjuan:2012wy}.
We present all these results in Fig.~\ref{fig:slopecomp}, where the smaller error is the statistical and the larger the quadratic combination of both statistical and systematic.
The comparison with the theoretical predictions mentioned in Sec.~\ref{intro} is also displayed.

Eventually, we want to comment on the effective single-pole mass determination $\Lambda_P$ from Eq.~(\ref{eq:tffpole}).
Using $b_P=m_P^2/\Lambda_{P}^2$ and the values in Eq.~(\ref{etaetapvalues}),
we obtain $\Lambda_{\eta}=0.706$ GeV and $\Lambda_{\eta^\prime}=0.833$ GeV. 
These values together with $\Lambda_{\pi}=0.750$ GeV obtained in Ref.~\cite{Masjuan:2012wy}
lead to $\Lambda_{\eta}<\Lambda_{\pi}<\Lambda_{\eta^\prime}$, in agreement with constituent-quark loops and VMD model approaches \cite{Ametller:1991jv}.

It is worth mentioning two interesting features of the low-energy parameters analysis performed above.
First, the values of the $\eta$ and $\eta^\prime$ time-like TFF at $q^2=-Q^2=112$ GeV$^2$ measured by the {\textit{BABAR}} Collaboration
\cite{Aubert:2006cy} do not modify our LEP determinations at the precision we are reporting in this work.
Second, and more important, given that the LEPs are defined at zero momentum transfer,
one would expect their fitted values to be dominated by low-energy data.
However, this is not the case: the high-energy data are relevant in order to reach higher PA sequences leading to more constrained values of the LEPs.
In the case at hand, only the {\textit{BABAR}} Collaboration provides precise measurements in the region between $5$ and $35$~GeV$^2$.
For instance, the value of the $\eta$ slope parameter shown in Eq.~(\ref{etaetapvalues}), $b_{\eta} = 0.60(6)(3)$,
turns out to be $b_{\eta} = 0.65(9)(7)$ when the {\textit{BABAR}} data are not included in the fits.
In view of this behavior and having in mind the $\pi^0$ TFF controversy after the measurements of the {\textit{BABAR}} \cite{Aubert:2009mc} and Belle \cite{Uehara:2012ag} Colls., a second experimental analysis by the Belle Coll.~covering this high-energy region would be very welcome.

Another interesting consequence of our analysis is the possible application of the present method to predict the time-like version of the TFFs.
Once the LEPs are fixed from a fit to experimental data in the space-like region,
the TFFs parametrized in the form of a given PA are well defined in the whole $Q^2$-complex plane except for possible genuine poles.
These poles are usually identified as resonances appearing in the time-like region, that is for $q^2=-Q^2>0$.
Therefore, the space-like TFF can be used as a suitable representation of the time-like TFF except in the vicinity of the poles attributed to resonances.
For the case of the $\eta$, the first possible vector resonance, the $\rho$ meson, is beyond the available phase-space.
Thus, one can take advantage of this fact and predict, for instance, the invariant spectrum of the $\eta\to\gamma e^+ e^-$ Dalitz decay
in a reliable and model-independent way.
In doing so, we find a nice agreement between our prediction and the reported experimental measurement \cite{Berghauser:2011zz}
(see also Ref.~\cite{Aguar-Bartolome:2013vpw} for a preliminary comparison of this prediction with more precise but not yet published experimental data
from the A2 Collaboration at MAMI).
For the case of the $\eta^\prime$, however, the $\rho$ and $\omega$ mesons are within the allowed kinematical region.
Because of that, the time-like TFF can only be described by the related space-like TFF in the low-energy region, far from these resonance poles, but not around them.

\section{{\boldmath$\eta$-$\eta^\prime$} mixing from the TFFs}
\label{sec:mixing}

In Sec. \ref{etaTFF}, the $\eta$ and $\eta^\prime$ TFFs were analyzed by means of the $P^N_{N}(Q^2)$ in order to predict the leading
$1/Q^2$ coefficients.
This information together with the predicted $F_{\eta^{(')}\gamma\gamma}(0)$, or their experimental measured values calculated from Eq.~(\ref{eq:tffdecay}), 
allows for the analysis of $\eta$-$\eta^\prime$ mixing.
This study can be performed either in the octet-singlet basis,
where the physical states are constructed employing the octet and singlet states,
or the quark-flavor basis, through the flavor states
$|\eta_q\rangle\equiv(|u\bar u\rangle+|d\bar d\rangle)/\sqrt{2}$ and $|\eta_s\rangle\equiv |s\bar s\rangle$.
In both cases, the leading $1/Q^2$ coefficients and the normalization of the TFFs at zero,
are written as functions of the different four pseudoscalar decay constants, defined as
$\langle 0|A^{(a,i)}_\mu|\eta^{(\prime)}(p)\rangle=i\sqrt{2}F^{(a,i)}_{\eta^{(\prime)}}p_\mu$,
where $a=8,0$ or $i=q,s$ depending on the chosen basis\footnote{
The axial-vector currents are defined as
$A^a_\mu=\bar q\gamma_\mu\gamma_5\frac{\lambda_a}{\sqrt{2}}q$, with
$A^q_\mu=\frac{1}{\sqrt{2}}(\bar u\gamma_\mu\gamma_5 u+\bar d\gamma_\mu\gamma_5 d)=
\frac{1}{\sqrt{3}}(A^8_\mu+\sqrt{2}A^0_\mu)$ and $A^s_\mu=\bar s\gamma_\mu\gamma_5 s=
\frac{1}{\sqrt{3}}(A^0_\mu-\sqrt{2}A^8_\mu)$.}.
For the reason explained below, we analyze $\eta$-$\eta^\prime$ mixing using the quark-flavor basis
\cite{Bramon:1997mf,Bramon:1997va,Feldmann:1998vh,Feldmann:1998sh,Feldmann:1999uf,Bramon:2000fr,Aloisio:2002vm,Escribano:2005qq,Ambrosino:2006gk,
Escribano:2007cd,Thomas:2007uy,Escribano:2008rq,Ambrosino:2009sc}. 
In this basis, the $\eta$ and $\eta^\prime$ decay constants are parametrized as
\begin{equation}
   \left(
      \begin{array}{cc}
          F^{q}_{\eta} & F^{s}_{\eta}\\
          F^{q}_{\eta^\prime} & F^{s}_{\eta^\prime}\\
      \end{array}
   \right)
=
   \left(
      \begin{array}{cc}
          F_{q}\cos\phi_{q} & -F_{s}\sin\phi_{s}\\
          F_{q}\sin\phi_{q} & F_{s}\cos\phi_{s}\\
      \end{array}
   \right)\ ,
\end{equation}
where $F_{q,s}$ are the light-quark and strange pseudoscalar decay constants, respectively,
and $\phi_{q,s}$  the related mixing angles.
Several phenomenological analyses find $\phi_q\simeq\phi_s$, 
which is also supported by large-$N_c$ ChPT calculations
where the difference between these two angles is seen to be proportional to an OZI-rule violating parameter and hence small \cite{Leutwyler:1997yr,Escribano:2005qq}.
This assumption, $\phi_q=\phi_s\equiv\phi$, is also a requirement of the FKS scheme \cite{Feldmann:1998vh,Feldmann:1999uf}.

Within this approximation, the asymptotic limits of the TFFs take the form
\begin{equation}
\label{eq:FLAVEtaAsymp}
\begin{split}
\lim_{Q^2\to\infty }Q^2F_{\eta\gamma^*\gamma}(Q^2) &=2(\hat c_q F^q_{\eta}+\hat c_s F^s_{\eta})\\
&=2(\hat c_q F_q\cos\phi-\hat c_s F_s\sin\phi)\ ,\\[1ex]
\lim_{Q^2\to\infty }Q^2F_{\eta^\prime\gamma^*\gamma}(Q^2) &=2(\hat c_q F^q_{\eta^\prime}+\hat c_s F^s_{\eta^\prime})\\
&=2(\hat c_q F_q\sin\phi+\hat c_s F_s\cos\phi)\ ,
\end{split}
\end{equation}
and their normalization at zero
\begin{equation}
\label{eq:FLAVEtaDecay}
\begin{split}
F_{\eta\gamma\gamma}(0) &=
\frac{1}{4\pi^2}\left(\frac{\hat c_q F^s_{\eta^\prime}-\hat c_s F^q_{\eta^\prime}}{F^s_{\eta^\prime}F^q_{\eta}-F^q_{\eta^\prime}F^s_{\eta}}\right)\\
&=\frac{1}{4\pi^2}\left(\frac{\hat c_q}{F_q}\cos\phi-\frac{\hat c_s}{F_s}\sin\phi\right)\ ,\\[1ex]
F_{\eta^\prime\gamma\gamma}(0) &=
\frac{1}{4\pi^2}\left(\frac{\hat c_q F^s_{\eta}-\hat c_s F^q_{\eta}}{F^s_{\eta}F^q_{\eta^\prime}-F^q_{\eta}F^s_{\eta^\prime}}\right)\\
&=\frac{1}{4\pi^2}\left(\frac{\hat c_q}{F_q}\sin\phi+\frac{\hat c_s}{F_s}\cos\phi\right)\ ,
\end{split}
\end{equation}
with $\hat c_q=5/3$ and $\hat c_s=\sqrt{2}/3$.

Using Eqs.~(\ref{eq:FLAVEtaAsymp}) and (\ref{eq:FLAVEtaDecay}), 
one can attempt to predict the mixing parameters in the quark-flavor basis, that is, the two decay constants, $F_q$ and $F_s$, 
and the single mixing angle $\phi$, with the results obtained in our fits.
However, only three of the four equations are independent, so, we have to choose the set of three equations that will be used to get 
the three mixing parameters.
Our choice is based on the precision achieved by the PAs.
While for the $\eta^\prime$ TFF the $P^N_{N}(Q^2)$ sequence reaches only the $N=1$ element,
with the consequent lack of stability checks and big uncertainties discussed above,
the $\eta$ TFF reaches $N=2$ (when the measured two-photon partial widths are included in the fits),
where the stabilization is attained and the uncertainty of the fitted parameters reduced.
Accordingly, we do not recommend to use the asymptotic limit of the $\eta^\prime$ TFF to extract the mixing parameters.
For the same reason, confident results for these parameters will be only obtained in the case of including the two-photon partial widths in the fits.
Nevertheless, for the sake of comparison, we will explore all the different possibilities for extracting such parameters.

We start considering our best scenario in terms of confidence and precision.
For the normalization at zero of both TFFs we use
$|F_{\eta\gamma\gamma}(0)|_{\rm exp}=0.274(5)$ GeV$^{-1}$ and $|F_{\eta^\prime\gamma\gamma}(0)|_{\rm exp}=0.344(6)$ GeV$^{-1}$
from the measured decay widths
$\Gamma_{\eta\to\gamma\gamma}=0.516(18)$ keV and $\Gamma_{\eta^\prime\to\gamma\gamma}=4.35(14)$ keV, respectively,
and for the asymptotic value of the $\eta$ TFF we take the value shown in Eq.~(\ref{mixinglimits}),
$\lim_{Q^2\to\infty }Q^2F_{\eta\gamma^*\gamma}(Q^2)=0.160(24)$ GeV.
With these values, the mixing parameters are predicted to be
\begin{equation}
\label{eq:mixresults}
\begin{split}
F_q/F_\pi&=1.06(1)\ ,\quad
F_s/F_\pi=1.56(24)\ ,\\[1.5ex]
&\qquad\phi=40.3(1.8)^{\circ}\ ,
\end{split}
\end{equation}
with $F_\pi=92.21(14)$ MeV \cite{PDG2012}.
These values represent a second important result of this work.
They can be compared, for instance, with the determination of the mixing parameters obtained in Ref.~\cite{Escribano:2005qq},
$F_q/F_\pi=1.10(3)$, $F_s/F_\pi=1.66(6)$ and $\phi=40.6(0.9)^{\circ}$,
after a careful analysis of $V\to\eta^{(\prime)}\gamma$, $\eta^{(\prime)}\to V\gamma$, with $V=\rho, \omega, \phi$, and
$\eta^{(\prime)}\to\gamma\gamma$ decays, and the ratio $R_{J/\psi}\equiv\Gamma(J/\psi\to\eta^\prime\gamma)/\Gamma(J/\psi\to\eta\gamma)$.
An update of the former values taking into account the latest experimental measurements of these decays gives
$F_q/F_\pi=1.07(1)$, $F_s/F_\pi=1.63(3)$ and $\phi=39.6(0.4)^{\circ}$.
An older phenomenological analysis based on the FKS scheme leads to
$F_q/F_\pi=1.07(3)$, $F_s/F_\pi=1.34(6)$ and $\phi=39.3(1.0)^{\circ}$ \cite{Feldmann:1998vh}
(see Ref.~\cite{Feldmann:1999uf} for a compendium of different results).
The agreement between these determinations and the values in Eq.~(\ref{eq:mixresults}) is quite impressive
since we only use the information of the TFFs to predict the mixing parameters.

If instead of using the asymptotic value of the $\eta$ TFF for the study of $\eta$-$\eta^\prime$ mixing,
we use the asymptotic value of the $\eta^\prime$ TFF in Eq.~(\ref{mixinglimits}),
the following results are found
\begin{equation}
\label{eq:mixresults2}
\begin{split}
F_q/F_\pi&=1.09(2)\ ,\quad
F_s/F_\pi=0.96(4)\ ,\\[1.5ex]
&\qquad\phi=33.5(0.9)^{\circ}\ ,
\end{split}
\end{equation}
in clear disagreement with all the values reported by the phenomenological analyses mentioned above and the results in
Eq.~(\ref{eq:mixresults}).
This discrepancy may be an indication of the lack of stability of the $P^1_1(Q^2)$ to predict the asymptotic limit.
However, the value we have obtained, $\lim_{Q^2\to\infty }Q^2F_{\eta^\prime\gamma^*\gamma}(Q^2)=0.255(4)$GeV,
is in accord with the \textit{BABAR} measurement in the timelike region at $q^2=112$ GeV$^2$,
$q^2|F_{\eta^\prime\gamma^*\gamma}(q^2)|_{112\, {\rm GeV}^2}=0.251(21)$GeV \cite{Aubert:2006cy}.
This contrasts with the situation for the $\eta$ TFF.
Our fitted value, $\lim_{Q^2\to\infty }Q^2F_{\eta\gamma^*\gamma}(Q^2)=0.160(24)$,
which we have used to get a reasonable estimate of the mixing parameters in Eq.~(\ref{eq:mixresults}),
is not in line with the value $q^2|F_{\eta\gamma^*\gamma}(q^2)|_{112\, {\rm GeV}^2}=0.229(31)$GeV
reported by the \textit{BABAR} Collaboration.
Given this situation, it might be the case that the mixing scheme used here is not complete enough
to catch the physical features of the $\eta$-$\eta^\prime$ mixing
(higher order effects of the chiral and large-$N_c$ expansions or including gluonium effects could be of certain relevance).
Therefore, precise determinations of the mixing parameters from lattice QCD techniques will be very welcome\footnote{
Recently, the ETM Collaboration has reported a value for the $\eta$-$\eta^\prime$ mixing angle in the quark flavor basis of
$\phi=46(1)_{\rm stat}(3)_{\rm sys}^\circ$ \cite{Michael:2013gka},
in good agreement with other lattice determinations,
$\phi=40.6(2.8)^\circ$, from the RBC and UKQCD Collaborations \cite{Christ:2010dd}, and
$\phi=42(1)^\circ$, from the Hadron Spectrum Collaboration \cite{Dudek:2011tt}.},
also for the implications of such mixing in the light-by-light scattering contribution to the anomalous magnetic moment of the muon,
which are the subject of the next section.

For completeness, we also provide two predictions:
the mixing parameters when the two-photon partial widths measurements are not included in the fits
and the asymptotic form factors when the updated values of the mixing parameters mentioned before are used.
For the first prediction, we take the asymptotic value of the $\eta$ TFF, obtained now with a $P^1_1(Q^2)$, i.e., $\lim_{Q^2\to\infty }Q^2F_{\eta\gamma^*\gamma}(Q^2)=0.168(10)$, 
as well as the predicted normalizations at zero,
$F_{\eta\gamma\gamma}(0)|_{\rm fit}=0.235(53)$ GeV$^{-1}$ and
$F_{\eta^\prime\gamma\gamma}(0)|_{\rm fit}=0.339(17)$ GeV$^{-1}$,
from Table \ref{tab:psresults}.
We find
$F_q/F_\pi=1.1(1)$, $F_s/F_\pi=1.5(2)$ and $\phi=43(5)^{\circ}$,
in fair agreement with the results in Eq.~(\ref{eq:mixresults}) but less precise.
For the second, we obtain from Eq.~(\ref{eq:FLAVEtaAsymp}) the values
$\lim_{Q^2\to\infty }Q^2F_{\eta\gamma^*\gamma}(Q^2)=0.163(4)$GeV and $\lim_{Q^2\to\infty }Q^2F_{\eta^\prime\gamma^*\gamma}(Q^2)=0.319(3)$GeV,
respectively, to be compared with the results shown in Eq.~(\ref{mixinglimits}).
We emphasize once more the impact of including the two-photon partial widths in the fits and
the relevance of getting higher order Pad\'e sequences in order to reach stability and reduce the uncertainties of the fitted parameters.
The latter, as discussed above, is not the case of the $\eta^\prime$ where a $P^1_1(Q^2)$ is the highest diagonal Pad\'e accessible for the fit.

\section{Implications on the hadronic light-by-light contribution to the \boldmath $(g-2)_{\mu}$}
\label{sec:g2}

The hadronic contributions to the anomalous magnetic moment of the muon $a_{\mu}$ consists on hadronic vacuum polarization as well as hadronic light-by-light scattering (HLBL).
The latter cannot be directly related to any measurable cross section and requires the knowledge of QCD contributions at all energy scales.
Since this is not known yet, one relies on hadronic models to compute it
\cite{Bijnens:1995xf,Hayakawa:1995ps,Bijnens:2001cq,Hayakawa:1997rq,Knecht:2001qf,Knecht:2001qg,Melnikov:2003xd,Dorokhov:2008pw,Hong:2009zw,Prades:2009tw,
Cappiello:2010uy,Greynat:2012ww,Goecke:2012qm,Masjuan:2012qn,Bijnens:2012an}.
Indeed, the theoretical value of $a_{\mu}$ is currently limited by uncertainties from the HLBL scattering contribution leading to an uncertainty in
$a_{\mu}$ of (2.6--4.0)$\times 10^{-10}$~\cite{Miller:2007kk,Jegerlehner:2009ry,Nyffeler:2009tw}
as well as the one from hadronic vacuum polarization (4.2--4.9)$\times 10^{-10}$ \cite{Davier:2010nc,Hagiwara:2011af}.

The present world average experimental value is given by $a_{\mu}^{\rm exp}=116 592 08.9(6.3)\times10^{-10}$~\cite{Bennett:2004pv,Bennett:2006fi},
still limited by statistical errors, and a proposal to measure the muon $(g-2)_{\mu}$ to a precision of $1.6 \times 10^{-10}$ has recently been submitted to FNAL
\cite{Carey:2009zzb}.
In view of this proposal, it is important to have better control on the HLBL contribution which as we will see may demand also better control on the TFF studied so far.


Using the large-$N_c$ limit of QCD \cite{tHooft:1973jz,Witten:1979kh} and also the chiral counting, it was proposed in \cite{deRafael:1993za}
to split the HLBL into a set of different contributions where the numerically dominant one arises from the pseudo-scalar exchange piece, the $a_{\mu}^{\rm HLBL,PS}$
(see Refs.~\cite{Knecht:2001qf,Jegerlehner:2009ry} for details):
$a_{\mu}^{{\rm HLBL},\pi^0}\sim 7\times 10^{-10}$ and $a_{\mu}^{{\rm HLBL},\eta^{(\prime)}}\sim 1.5\times 10^{-10}$.
The main ingredient on the determination of the pseudoscalar-exchange process $a_{\mu}^{\rm HLBL,PS}$ is the double off-shell TFF
$F_{P^*\gamma^*\gamma^*}((q_1+q_2)^2,q_1^2, q_2^2)$ dominated by an on-shell pseudoscalar \cite{Knecht:2001qf}.
The TFF should be considered to be off shell
(see Refs.~\cite{Jegerlehner:2009ry,Melnikov:2003xd,Dorokhov:2008pw,Nyffeler:2009tw,Cappiello:2010uy} where this point is addressed).
Since such effects for the $\eta$ and $\eta^\prime$, which are expected to be small, are not known, we should keep the pseudoscalar-pole simplifications in our calculations.
A preliminary discussion on off-shell effects is reported below.

In this section we plan to study the impact of the results obtained in Sec. \ref{sec:PSTFF} to the HLBL with the intuition that it is more important
to have a good description at small and intermediate energies, e.g., by reproducing the slope and curvature of the TFFs,
than a detailed short-distance analysis since the angular integrals used to compute $a_{\mu}^{\rm HLBL}$ do not seem to be very sensitive to
the correct asymptotic behavior for large momenta \cite{Knecht:2001qf}.


In the large-$N_c$ limit, QCD Green's functions are meromorphic functions with simple poles and no branch cuts since consist of
infinitely many noninteracting sharp mesons states whose masses and decay constants are in principle unknown. 
As such sum is not known in practice, one ends up truncating the spectral function in a resonance saturation scheme, the so-called minimal hadronic approximation
\cite{Peris:1998nj}.
The resonance masses used in each calculation are then taken as the physical ones from PDG instead of the corresponding masses in the large-$N_c$.
This assumption together with the effect of the spectrum truncation should be taken into account on the final systematic error \cite{Masjuan:2007ay,Queralt:2010sv}.

A way of evade these caveats comes from the Pad\'e theory \cite{Masjuan:2007ay}. 
In this context, one defines the TFF as a PA defined from its LEPs \cite{Masjuan:2012qn}:
\begin{equation}
\label{TFFP01}
F_{P\gamma^*\gamma^*}^{P01}(Q_1^2,Q_2^2) = P_1^0(Q_1^2,Q_2^2)\,=a \frac{b}{Q_1^2+b}\frac{b}{Q_2^2+b}\ ,
\end{equation}
where the free parameters are matched at low energies with the results in Table \ref{tab:psresults2},
$a$ is fixed by $\Gamma_{P\rightarrow \gamma\gamma}$ and $b$ by the slope $b_{P}$.

The convergence of the PA sequence to a meromorphic function is guaranteed by Pomerenke's theorem \cite{Pommerenke}.
The problem is to know how fast this convergence is and also how to ascribe a systematic error on each element of that sequence.
For meromorphic functions, the simplest way of evaluating a systematic error is by comparing the difference between two consecutive elements on the PA sequence
\cite{Masjuan:2009wy} (see Appendix~\ref{App} for details).

In our approach to the TFF, the second element on the PA sequence is given by:
\begin{equation}\label{TFFP12}
\begin{split}
&F_{P\gamma^*\gamma^*}^{P12}(Q_1^2,Q_2^2) =\\
&= P_2^1(Q_1^2,Q_2^2)\,=\frac{a+b\,Q_1^2}{(Q_1^2+c)(Q_1^2+d)}\frac{a+b\, Q_2^2}{(Q_2^2+c)(Q_2^2+d)}\ ,
\end{split}
\end{equation}
with four coefficients to be matched with $\Gamma_{P\rightarrow \gamma\gamma}$, the slope $b_{P}$, the curvature of the TFF $c_{P}$ and
the first effective vector meson resonance accounted for the appropriate $\rho,\omega,\phi$ mixing \cite{Landsberg:1986fd}, illustrated in Fig.~\ref{fig:PL1poles}.
The error for the effective mass is taken from the half-width rule \cite{Masjuan:2012gc,Masjuan:2012sk}.
The results are collected in Table \ref{tab:g2}. 

The weighted average results for the low-energy parameters of the $\eta^\prime$-TFF collected in Table~\ref{tab:psresults2}
considered only the $P^L_1(Q^2)$ sequence since with the $P^N_N(Q^2)$ we only reached its first element.
Therefore, in the determination of the $a_{\mu}^{{\rm HLBL},\eta'}$ in Eq.~(\ref{TFFP12}) we used the low-energy parameter of order ${\cal O} (Q^2 )^3$
instead of the effective mass obtained in Ref.~\cite{Landsberg:1986fd}.
Both procedures yield very similar results.

\begin{table*}[ht]
\centering
\begin{tabular}{ | c | c | c | c | c | }
\hline
$F_{P\gamma\gamma^*}(Q_1^2,Q_2^2)$ &       $\eta$    & $\eta^\prime$    &  Total            \\ [4pt] \hline \hline
$P_1^0(Q_1^2,Q_2^2)$                  &\hspace{0.2cm} $1.25(15)$ \hspace{0.2cm}  &\hspace{0.2cm} $1.21(12)  $ \hspace{0.2cm} & $8.96(59)$  \\ [3pt] \hline 
$P_2^1(Q_1^2,Q_2^2)$                  & $1.27(19)$   & $ 1.22(12) $  & $ 9.00(74) $  \\ [3pt]\hline \hline
Eq.~(\ref{HW1})                             & $1.44(19)$   & $ 1.27(29) $  & \hspace{0.2cm}  $8.84(35)$ \hspace{0.2cm}   \\ [3pt]\hline
Eq.~(\ref{HW2})                            & $1.38(16)$   & $ 1.22(9) $  & $ 8.48(45) $  \\ [3pt]\hline
\end{tabular}
\caption{Collection of results for the $a_{\mu}^{\rm HLBL,PS}$ for ${\rm PS}=\eta$ and $\eta^\prime$ contributions.
The last column contains also the result obtained in Ref.~\cite{Masjuan:2012qn} for the $\pi^0$-TFF, with errors combined in quadrature.
Results in units of $10^{-10}$.}
\label{tab:g2}
\end{table*}

The similarity of the results obtained with both approximants (\ref{TFFP01}) and (\ref{TFFP12}) indicates, as expected~\cite{Bijnens:2012an},
that the low-energy region (up to 1--2 GeV) dominates the contribution to $\amu^{\rm HLBL,PS}$.
To evaluate the error on our approximation we look for the maximum of the difference in the region up to $1$~GeV between
the $P^0_1(Q_1^2,Q_2^2)$ and $P^1_2(Q_1^2,Q_2^2)$ as explained in Ref.~\cite{Masjuan:2009wy}.
Of course, this difference depends on the energy, and grows as the energy increases.
At $1$~GeV, the relative difference is about $5\%$, and we take this error as a conservative estimate of the error on the whole low-energy region.
We should add this error to the $\amu^{\rm HLBL,PS}$ final result.
In Appendix~\ref{App}, a more rigorous way of estimating such error is presented.
 
In order to provide $a_{\mu}^{\rm HLBL,PS}$, we also collect the results for the $a_{\mu}^{{\rm HLBL},\pi^0}$ obtained in Ref.~\cite{Masjuan:2012qn},
where the same method was used but with the full off-shell TFF, i.e, $a_{\mu}^{{\rm HLBL},\pi^0}=6.49(56)\cdot 10^{-10}$ and 
$a_{\mu}^{{\rm HLBL},\pi^0}=6.51(71)\cdot 10^{-10}$ corresponding to the first and second elements on the PA sequence, respectively.

There is a second way of computing the HLBL, which incorporates $1/N_c$ corrections,
that would reassess our previous results. In this way, one makes use of the meson dominance and the half-width rule when accounting for the TFF
(see Refs.~\cite{RuizArriola:2010fj,Masjuan:2012gc,Masjuan:2012sk,Arriola:2012vk}).
Meson dominance means to take the high-energy behavior given by pQCD and the minimal number of mesons to satisfy its condition \cite{Masjuan:2012sk,Arriola:2012vk}.
Then, errors in the meson-dominated form factors are estimated by the half-width rule, i.e.,
by treating resonance masses as random variables distributed with the dispersion given by its decay width. In this way, 

\begin{eqnarray}\label{HW1}
F_{P\gamma^*\gamma}(Q^2)= \frac1{4 \pi^2 F_{P}}\frac{m_{\rm eff}^2}{m_{\rm eff}^2+Q^2}\ ,
\end{eqnarray}
provided one has the relation $m_{\rm eff}^2=8\pi^2F_{P}(\hat c_q F^q_P+\hat c_s F^s_P)$ for $P=\eta,\eta^\prime$ to satisfy the asymptotic limit~(\ref{eq:FLAVEtaAsymp}). Numerical evaluations are performed with the results in Eq.~(\ref{mixinglimits}). For the $\pi^0$ case, $m_{\rm eff}^2=8\pi^2F_{\pi}^2$. 

If we improve our model by including a second resonance, $m_{\rm eff}$ and $m_{\rm eff'}$, we get, after imposing the anomaly and large-$Q^2$ behavior, 
\begin{eqnarray}\label{HW2}
\begin{split}
\hspace{-0.7cm}F_{P\gamma^*\gamma}(Q^2)= &\\
= \frac1{4 \pi^2 F_{P}}  &\frac{m_{\rm eff}^2 m_{\rm eff'}^2 + 8 \pi^2  F_{P} (\hat c_q F^q_{P}+\hat c_s F^s_{P}) Q^2}
{(m_{\rm eff}^2+Q^2)(m_{\rm eff'}^2+Q^2)}\ , \label{eq:2m}
\end{split}
\end{eqnarray}
\noindent
where $m_{\rm eff}$ and $m_{\rm eff'}$ correspond to the VMD for each $\pi^0,\eta,\eta^\prime$ TFF \cite{Landsberg:1986fd}:
for the $\eta$ we obtain $m_{\rm eff}=0.732(71)$ GeV and for the $\eta^\prime$ $m_{\rm eff}=0.822(58)$ GeV, as described in Sec.~\ref{etaTFF};
we also obtain $m_{\rm eff'}=1.41(21)$~GeV and $m_{\rm eff'}=1.51(16)$~GeV for $\eta$ and $\eta'$, respectively
(using $m_{\omega'}=1.425$~GeV, $\Gamma_{\omega'}=0.215$~GeV, $m_{\phi'}=1.680$~GeV, and $\Gamma_{\phi'}=0.150$~GeV).
The errors come from the half-width rule.
Numerical evaluations are performed again with the results in Eq.~(\ref{mixinglimits}).
For the $\pi^0$ case, $m_{\rm eff}=m_{\rho}$, $m_{\rm eff'}=m_{\rho'}$, and the asymptotic limit is $2F_{\pi}$.  

The results using both Eqs.~(\ref{HW1}) and (\ref{HW2}) are shown in Table \ref{tab:g2}.
For the $\pi^0$ case, using $m_\rho=0.775$~GeV, $m_{\rho'}=1.465$~GeV, $\Gamma_\rho=0.148$~GeV and $\Gamma_{\rho'}=0.400$~GeV,
we obtain $\amu^{{\rm HLBL},\pi^0}=6.13(1)\cdot 10^{-10}$ and $\amu^{{\rm HLBL},\pi^0}=5.88(41)\cdot 10^{-10}$ using Eqs.~(\ref{HW1}) and (\ref{HW2}), respectively. Equation~(\ref{HW2}) yields always smaller results due to the fact that its slope is always larger ($b_P =  m_{\rm P}^2 \sum 1/m_{\rm eff}^2$) than the one from Eq.~(\ref{HW1}) ($b_P =  m_{\rm P}^2/m_{\rm eff}^2$). The difference between the results using both equations should be accounted for by an extra source of systematic error.

The nice agreement between all the different determinations of $a_{\mu}^{\rm HLBL,PS}$ collected in Table \ref{tab:g2} is quite reassuring.
We should take the result using the $P_1^0(Q_1^2,Q_2^2)$  as our main result and the others as a cross-check of that one.
On top we should add a systematic error of about $5\%$ yielding our final number as:

\begin{equation}\label{g2result}
a_{\mu}^{\rm HLBL,PS} = 9.0(6)(4)\times 10^{-10}\ ,
\end{equation}
where the first error comes from the input errors and the second from the systematic error.

This result, which for the first time contains a systematic error, is in nice agreement with most of the recent phenomenological calculations for such quantity
\cite{Prades:2009tw}, but smaller than those determinations that model an off-shell TFF
\cite{Jegerlehner:2009ry,Melnikov:2003xd,Dorokhov:2008pw,Nyffeler:2009tw},
pointing towards a positive impact of the off-shellness of the pseudoscalar in such computation. Indeed, assuming $U(3)$ and chiral symmetries
(neglecting the effect of nonzero quark masses, the $\eta-\eta'$ mixing and the possible gluonic contribution to the axial anomaly coming from the $\eta'$),
one can parameterize the off-shellness of the $\eta$ and $\eta'$ in the TFF through the quark condensate magnetic susceptibility $\chi$
as done for the $\pi^0$ contribution in Refs.~\cite{Nyffeler:2009tw,Jegerlehner:2009ry,Masjuan:2012qn}.
In such a way one can promote the TFF in Eqs.~(\ref{TFFP01}) and (\ref{TFFP12}) to their full off-shell counterparts in Ref.~\cite{Masjuan:2012qn}.
One would obtain, then, $1.51(23) \times 10^{-10}$ and $1.91(35) \times 10^{-10}$ for $a_{\mu}^{{\rm HLBL},\eta(\eta')}$ respectively.
The error is a quadratic combination of the error report in Table \ref{tab:g2} together with the error coming from the magnetic susceptibility $\chi$
(which leads to the errors $0.17\times10^{-10}$ and $0.33 \times 10^{-10}$ for the $\eta$ and the $\eta'$ contributions, respectively).
While on-shell and off-shell TFFs yield compatible result for the $\eta$ case, the same is not true for the $\eta'$ and the compilation of pseudoscalar contributions
in this scenario would yield $a_{\mu}^{\rm HLBL,PS} = 9.9(7)(5)\times 10^{-10}$, one standard deviation larger than the result in Eq.~(\ref{g2result}).
This exercise provides an insight on the role of off-shell effects in the pseudoscalar exchange contribution although one should not take them as definitive since,
for example, $U(3)$ breaking effects (i.e, the difference between $m_{\eta}$ and $m_{\eta'}$, its mixing or its gluonic content) may be important.
Reference~\cite{Dorokhov:2011zf} found, however, opposite conclusions, with a negative impact of the off-shellness of the pseudoscalars within the nonlocal quark model.

A different approach to this problem may come from the study of the light- and strange-quark contributions to the HLBL instead of the ones from the
$\pi^0, \eta$ and $\eta'$.
Recent progress from the lattice collaborations to simulate the two-photon partial decay width from both light- and strange-quark components suggest
the viability of such an approach\footnote{
We thank C. Urbach for discussions and correspondence along these lines.}.
These results together with its corresponding form factors with one and two virtual photons can be used to cross-check our results and assumptions.

Reference~\cite{Prades:2009tw} provides a compilation of the different contributions to the HLBL.
They report a final value $a_{\mu}^{\rm HLBL}=(10.5\pm2.6)\cdot 10^{-10}$.
Updating the pseudoscalar piece of the HLBL considered in Ref.~\cite{Prades:2009tw} using Eq.~(\ref{g2result}), we obtain $a_{\mu}^{\rm HLBL}=(8.1\pm2.4)\cdot 10^{-10}$. This shift implies that the difference $\Delta a_{\mu} = a_{\mu}^{\rm exp} - a_{\mu}^{\rm SM}$ grows from $3.6$ standard deviations ~\cite{Jegerlehner:2009ry} to $3.9$ standard deviations, showing the role of precise information on TFF to constrain the HLBL piece of the muon $(g-2)$.

Primakoff determination of the two-photon partial decay width of the $\eta$ meson is not included in the averaged fit on the Particle Data Tables \cite{PDG2012}.
If one would use that result (i.e., $\Gamma_{\eta \rightarrow \gamma \gamma}=476(63)$~eV) instead, the result for the $\amu^{{\rm HLBL},\eta}$ would be reduced by $7\%$.
We remark, then, the need of a precise measurement for such partial decays in order to better constraint the impact of the TFF in the HLBL.
We notice that such determination with precision of about $1\%$ would imply an error on HLBL of the same order.

\section{Conclusions}
\label{sec:conc}

The experimental data on the $\eta$ and $\eta^\prime$ transition form factors in the space-like region have been analysed
at low and intermediate energies in a model-independent way through the use of rational approximants,
thus extending and complementing the previous work done for the $\pi^0$ case.
The method of Pad\'e approximants is simple, systematic and provides an estimate of the systematic errors.
The slope and curvature parameters of the form factors as well as their values at zero and infinity have been extracted.
The slopes of both pseudoscalar mesons are well within phenomenological determinations although these,
based on the VMD approach, do not include a systematic error associated to the model dependence.
The curvatures are presented for the first time. 
At the current level of accuracy, they follow in both cases the VMD prescription $c_P=b_P^2$.
However, VMD should be taken only as the simplest model-dependent approximation to a more general model-independent rational parametrization, which at present is still compatible with data.
The normalization of the form factors, when this information is not used as input data,
is seen to be consistent with measurements, even though less precise.
The incorporation of these measurements into the fits reduces drastically the uncertainty on the $\eta$ low-energy parameters,
besides reducing the systematic errors, not the case of the $\eta^\prime$, where they are not affected by this inclusion.
The asymptotic behavior of the form factors is in nice agreement with the \textit{BABAR} reported values at $q^2=112$ GeV$^2$,
better in the case of the $\eta^\prime$ than the $\eta$.
The influence of our results on the mixing parameters of the $\eta$-$\eta^\prime$ system has been also discussed.
The values obtained in the quark-flavor basis are in accord with existing phenomenological analyses only in the case of 
employing the prediction of the asymptotic value of the $\eta$ transition form factor in the extraction of those parameters.
Finally, making use of the Pad\'e techniques and the large-$N_c$ results obtained in Ref.~\cite{Knecht:2001qf},
we have shown the impact of our investigations in the determination of the pseudoscalar-exchange contributions
to the hadronic light-by-light scattering part of the anomalous magnetic moment $a_\mu$.

\acknowledgements

We thank  C.~F.~Redmer, M.~Unverzagt, C.~Urbach, and M.~Vanderhaeghen  for discussions.
We also thank S.~Eidelman and S.~Peris for a careful reading of the manuscript.
Work was supported by the Deutsche Forschungsgemeinschaft DFG through the Collaborative Research Center 
``The Low-Energy Frontier of the Standard Model" (SFB 1044),
and partially by MICINN of Spain (FPA2010-16696), Junta de Andaluc\'ia (Grants No. P07-FQM 330 and No. P08- FQM 101).
This work was also supported in part by the Ministerio de Ciencia e Innovaci\'on under grants FPA2011-25948 and No. AIC-D-2011-0818,
the European Commission under the 7th Framework Program through the``Research Infrastructures'' action of the
``Capacities'' Programme Call: FP7-INFRA-STRUCTURES-2008-1 (Grant Agreement No. 227431),
the Spanish Consolider-Ingenio 2010 Program CPAN (CSD2007-00042),
and the Generalitat de Catalunya under Grant SGR2009-00894.

\appendix

\section{Parameterization of our best Pad\'e approximant fits}\label{AppTL}

In this Appendix we provide the parameterizations of our best $P^L_1(Q^2)$ fits for both the $\eta$- and $\eta^\prime$-TFFs.
Defining $P^L_1(Q^2)$ as 

\begin{equation}
P^L_1(Q^2) = \frac{T_L(Q^2)}{R_1(Q^2)} = \frac{t_1 Q^2+ t_2 Q^4+\cdots t_L (Q^2)^L}{1+r_1 Q^2}\ ,
\end{equation}
the corresponding fitted coefficients\footnote{
For full precision of the coefficients together with the correlation matrix, contact the corresponding author.}
for both $\eta$- and $\eta^\prime$-TFF are collected in Table \ref{tab:param}.
$L=5$ for the $\eta$ case and $L=6$ for the $\eta^\prime$. 

\begin{table*}[ht]
\centering
\begin{tabular}{ | c | c | c |}
\hline
 &	$\eta$-TFF	&	$\eta^\prime$-TFF \\ \hline
 \hspace{0.2cm}	$t_1$	\hspace{0.2cm}	&	\hspace{0.3cm} 	$0.274 $	\hspace{0.3cm}	&	\hspace{0.5cm}	$0.343$	\hspace{0.5cm}   \\ 
$t_2$	& $0.011 $ & $0.007 $ \\
$t_3$	& $0.789\cdot 10^{-3}$ & $0.986 \cdot 10^{-3}$  \\
$t_4$	& $0.229\cdot 10^{-4}$ & $0.744 \cdot 10^{-4}$  \\
$t_5$	& $0.169\cdot 10^{-6}$ & $0.252 \cdot 10^{-5}$  \\
$t_6$	& & $0.290 \cdot 10^{-7}$  \\ \hline
$r_1$	& $1.968$ & $1.442$  \\
\hline                                           
\end{tabular}
\caption{Fitted coefficients for our best $P^L_1(Q^2)$ for the $\eta$- and $\eta^\prime$-TFF.}
\label{tab:param}
\end{table*}

\section{Test of convergence of the PA sequence with a model}
\label{App}

To test how fast the convergence of our PA sequence is, we consider a Regge model for a pseudoscalar TFF 
(see, for example, \cite{Arriola:2006ii,Arriola:2010aq,Kampf:2011ty} where similar models are used to study the $\pi^0$-TFF).
For ease of manipulation, we construct the model in the large-$N_c$ limit ($N_c$ been the number of colors). 
In this limit, the vacuum sector of QCD becomes a theory of infinitely many noninteracting mesons and the propagators of the hadronic amplitudes
are saturated by infinitely many sharp meson states.
In the particular case below, the pseudoscalar couples first to a pair of vector mesons $V_{\rho,\omega,\phi}$ which then transform into photons.
Thus, we have

\begin{equation}
\begin{split}\label{Reggemodel1}
F_{P\gamma^*\gamma^*}(q^2_1,q^2_2)=&\\
\sum_{V=V_{\rho},V_{\omega},V_{\phi}}&\frac{F_{V}(q^2_1)F_{V}(q^2_2)G_{P V V}(q^2_1,q^2_2)}
{(q^2_1-M^2_{V})(q^2_2-M^2_{V})}\ ,
\end{split}
\end{equation}
\noindent
where $P=\eta,\eta'$, $F_{V}$ is the current-vector meson coupling and
$G_{PVV}$ is the coupling of two vector mesons to the pseudoscalar $\eta$ or $\eta^\prime$~\cite{Landsberg:1986fd}. 
The sum in Eq.~(\ref{Reggemodel1}) should run, for each vector channel ($\rho,\omega$ or $\phi$), over all its radial excitations.
The dependence on the resonance excitation number $n$ is the following

\begin{equation}
\begin{split}
M^2_{V_{\rho}}&=M^2_{V_{\omega}}=\frac{1}{a}M^2_{V_{\phi}}=M^2+ n \Lambda^2\, ,\\ \nonumber
F_{V_{\rho}}&=N_c {V_{\omega}} = -\frac{N_c}{\sqrt{2}}  {V_{\phi}} \equiv F\ .
\end{split}
\end{equation}
\noindent
with $a=1.3$~\cite{Landsberg:1986fd,Ametller:1991jv}.
The combination of sums in Eq.~(\ref{Reggemodel1}) can be expressed in terms of the Digamma function $\psi(z)=\frac{d}{dz}\log\Gamma(z)$:
\begin{widetext}
\begin{equation}\label{Reggemodel2}
F_{P\gamma^*\gamma^*}(q^2_1,q^2_2) =  F_{P\gamma^*\gamma^*}(Q^2,A)=
\frac{c}{N_c A Q^2} \left[ \psi\left(\frac{ M^2}{\Lambda^2}+\frac{Q^2(1+A)}{2 \Lambda^2}\right)-\psi\left(\frac{M^2}{\Lambda^2}+\frac{Q^2(1-A)}{2 \Lambda^2}\right)\right]\ ,
\end{equation}
\end{widetext}
\noindent
where $Q^2=-(q^2_1+q^2_2)$, $A=\frac{q^2_1-q^2_2}{q^2_1+q^2_2}$ and $c$ a constant~\cite{Arriola:2006ii,Arriola:2010aq}.

To reassemble the physical case we consider $N_c=3$, $\Lambda^2=1.3$~GeV$^2$
(as suggested by the recent light nonstrange $q\bar{q}$ meson spectrum analysis \cite{Masjuan:2012gc} using the half-width rule \cite{Masjuan:2012sk}),
$A=1/2$, $M^2= \lambda\times 0.64$ GeV$^2$ ($\lambda =0.95$ for the $\eta$ TFF and $\lambda =1.05$ for the $\eta^\prime$ TFF 
using the standard VMD scheme~\cite{Landsberg:1986fd}) and the constant $c$ in such a way that the anomaly $F_{P\gamma\gamma}(0,0)=1/(4\pi^2F_{P})$ is recovered.
In the following, we consider only the example for the $\eta$ TFF. 
The $\eta^\prime$ case is very similar and does not gives new information.

Once the model is defined, by generating a set of pseudodata we can test how fast the PA sequence converge to $Q^2 F_{P\gamma^*\gamma}(Q^2)$.
Considering 10 points in the region $0.6 < Q^2 <2.2 $~GeV$^2$, 15 points in the region $2.7 < Q^2 <7.6 $~GeV$^2$, and 10 more points in the region $8.9 < Q^2 <34 $~GeV$^2$, 
we are able to resemble the physical situation. We fit these pseudodata with a $P^L_1(Q^2)$ sequence and we collect the LEPs obtained with them (going up to $L=7$) in Table~\ref{tab:convergence}. 


The last column in Table~\ref{tab:convergence} shows the LEPs calculated from the model. 
Comparing each entry in this table with the corresponding value from the last column we can clearly see a pattern of convergence. 
For example, with a $P^4_1(Q^2)$ the slope and the curvature are determined with $6\%$ and $19\%$ of error, respectively. 
With a $P^6_1(Q^2)$, such errors reduce to $3\%$ and $10\%$, respectively. A similar study can be done with the $P^N_N(Q^2)$ sequence. In this case, with the $P^1_1(Q^2)$, slope and curvature are determined with $15\%$ and $50\%$ of error, respectively. With the $P^2_2(Q^2)$ the errors reduce drastically to $0.5\%$ and $2\%$, respectively. Since the uncertainty of the LEPs determination with the $P^1_1(Q^2)$ is much larger than with the $P^2_2(Q^2)$, the $P^1_1(Q^2)$ is never used in this work. Moreover, the errors for the LEPs from the  $P^2_2(Q^2)$ are even smaller than those from the  $P^5_1(Q^2)$, allowing a comparison among them.
Since our pseudodata have no errors, these determinations give us an idea of the genuine error done due to the fact that the PA sequence is finite, 
independently of the statistical errors in the data points. 
We call such kind of error a \emph{systematic} error. 
Such errors depend also on the amount of data points. Including more data points, especially in the low-energy region, diminishes all the systematic errors. 
This exercise is model dependent. 
In Ref.~\cite{Masjuan:2012wy} different models were analyzed with the purpose of obtaining a conservative systematic error for each LEPs at a given $L$, 
ascribing as a final systematic error a value around $5\%$ and $20\%$ for slope and curvature, respectively, for a $P^5_1(Q^2)$.
These are the results used in the present work. 

 \begin{table*}[ht]
\caption{Pattern of convergence for a $P^L_1(Q^2)$ sequence up to $L=6$ for the value $F_{\eta\gamma\gamma}(0,0)$ and the first two derivatives $b_{\eta}$ and $c_{\eta}$. The last column shows the exact results obtained with the model in Eq.~(\ref{Reggemodel2}).}
\begin{center}
\begin{tabular}{|c||c|c|c|c|c|c|c||c|}
\hline
 & $P^1_1(Q^2)$ &$P^2_1(Q^2)$ & $P^3_1(Q^2)$ & $P^4_1(Q^2)$ & $P^5_1(Q^2)$ & $P^6_1(Q^2)$ &$P^7_1(Q^2)$ & $F_{\eta\gamma^*\gamma}(Q^2,0)$ \\
\hline
$F_{\eta\gamma\gamma}(0,0)$ & $0.278$ & $0.276$ & $0.276$ & $0.276$ & $0.275$ & $0.275$ & $0.275$ & $0.275$   \\
$b_{\eta}$ & $0.492$ & $0.471$ & $0.458$ & $0.450$ & $0.442$ & $0.437$& $0.434$ & $0.426$ \\
$c_{\eta}$ & $0.242$ & $0.220$ & $0.205$ & $0.196$ & $0.188$ & $0.182$& $0.178$ & $0.166$  \\
\hline
\end{tabular}
\end{center}
\label{tab:convergence}
\end{table*}

\begin{figure}
\begin{center}
\includegraphics[width=8.0cm]{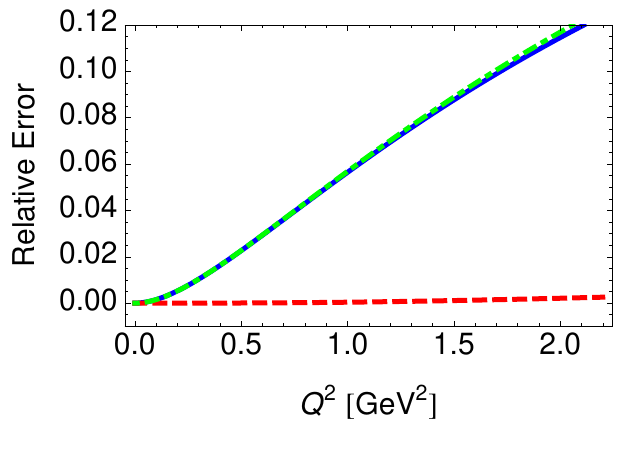}
\caption{Relative error for the first and second elements of the $P^N_{N+1}(Q^2)$ sequence compared to the TFF function Eq.~(\protect \ref{Reggemodel2})
(blue and red-dashed lines, respectively).
The green-dotdashed line represents the relative error between the first and the second element on the approximant sequence.}
\label{fig:syserror}
\end{center}
\end{figure}

Equations~(\ref{Reggemodel1}) and (\ref{Reggemodel2}) use the large-$N_c$ and chiral limits and thus have an analytic structure in the complex momentum plane
which consists of an infinity of isolated poles but no branch cut, i.e. they become meromorphic functions.
As such, they have a well-defined series expansion in powers of momentum around the origin with a finite radius of convergence given by the first resonance mass.
It is well known \cite{Pommerenke} and largely explored in the context of large $N_c$ \cite{Masjuan:2007ay,Masjuan:2008fr,Queralt:2010sv}
that the convergence of any near diagonal PA sequence to the original function for any finite momentum,
over the whole complex plane (except perhaps in a zero-area set) is guaranteed.

By expanding Eq.~(\ref{Reggemodel2}) one obtains the LEPs that are used to build up the $P^N_{N+1}(q^2_1,q^2_2)$ sequence.
Each element of this sequence approximates better the low-energy region than the intermediate or large one, although the larger the sequence,
the larger the region well approximated.
Comparing the $P^0_1(q^2_1,q^2_2)$ and the $P^1_2(q^2_1,q^2_2)$ with $F_{P\gamma^*\gamma^*}(q^2_1,q^2_2)$ from Eq.~(\ref{Reggemodel2})
one gets an idea of that $q^2$-dependent systematic error.
We show in Fig.~\ref{fig:syserror} the relative error for both $P^0_1(q^2_1,q^2_2)$ and $P^1_2(q^2_1,q^2_2)$ compared to $F_{\eta\gamma^*\gamma^*}(q^2_1,q^2_2)$
(blue and red-dashed lines) but also the relative error between $P^0_1(q^2_1,q^2_2)$ and $P^1_2(q^2_1,q^2_2)$ (green-dotdashed line).
We remark the similarity between the relative error of the $P^0_1(q^2_1,q^2_2)$ and the one between $P^0_1(q^2_1,q^2_2)$ and $P^1_2(q^2_1,q^2_2)$.
This simple exercise suggests to use such difference to estimate the systematic error done with the $P^0_1(q^2_1,q^2_2)$. 

In such a way, we define an error function $\epsilon(Q_1^2,Q_2^2))$ as
\begin{equation}\label{TFFe}
F_{P\gamma^*\gamma^*}(Q_1^2,Q_2^2) = P_1^0(Q_1^2,Q_2^2)(1+\epsilon(Q_1^2,Q_2^2))\ ,
\end{equation}
with $P_1^0(Q_1^2,Q_2^2)$ given in Eq.~(\ref{TFFP01}) and $\epsilon(Q_1^2,Q_2^2)$ emulating the difference between
$P^0_1(q^2_1,q^2_2)$ and $P^1_2(q^2_1,q^2_2)$.
As shown in Fig.~\ref{fig:syserror}, the error increases with the energy reaching almost $10\%$ of relative error for energies around $2$~GeV,
the region which dominates the $a_{\mu}^{{\rm HLBL}}$. 
The error function can be naively parameterized as:
\begin{equation}\label{error}
\epsilon(Q_1^2,Q_2^2))=\bigg(1+\frac{Q_1^2}{20} \bigg)\bigg(1+\frac{Q_2^2}{20} \bigg)\ .
\end{equation}

Computing the angular integrals accounting for $a_{\mu}^{{\rm HLBL}}$ with Eq.~(\ref{TFFP01}) or Eq.~(\ref{TFFe}) yields a difference around
$3\%$ for the $\eta$ case and around $5\%$ for the $\eta^\prime$ (larger due to the larger normalization of the TFF).
We suggested in the main text to ascribe $5\%$ of  systematic error for both $\eta$ and $\eta^\prime$ TFFs differences,
which should account for any possible model-dependent extraction of such error.


%

\end{document}